\documentclass[aps,prl,reprint,groupedaddress]{revtex4-1}

\pdfoutput=1
\usepackage{amsmath}
\usepackage{amssymb}
\usepackage{graphicx,mathrsfs}
\usepackage{nicefrac}

\newcommand{\be}{\begin{equation}}
\newcommand{\ee}{\end{equation}}
\newcommand{\bea}{\begin{eqnarray}}
\newcommand{\eea}{\end{eqnarray}}

\newcommand{\nn}{\nonumber}

\usepackage{color}

\begin{document}


\title{Measuring the diphoton coupling of a 750 GeV resonance}


\author{S.~Fichet}
\affiliation{ICTP-SAIFR, IFT, S\~ao Paulo State University, Brazil}

\author{G.~von Gersdorff}
\affiliation{Departamento de F\'isica, Pontif\'icia Universidade Cat\'olica de Rio de Janeiro, Rio de Janeiro, Brazil}

\author{C.~Royon}
  \affiliation{Kansas University, Lawrence, USA}
  \affiliation{Nuclear Physics Institute (PAN), Cracow, Poland}



\begin{abstract}

A slight excess  has been observed in the first data of photon-photon events  at the 13~TeV LHC, that might be interpreted as a  hint of physics beyond the Standard Model. 
We show that a completely model-independent measurement of the photon-photon coupling of a putative 750 GeV resonance will be possible  using the forward proton detectors scheduled at ATLAS and CMS.   

\end{abstract}


\maketitle

\section{Introduction}

The Large Hadron Collider (LHC) is currently performing collisions at the unprecedented center-of-mass energy of $13$~TeV.
Its primary goal  is the search for physics beyond the Standard Model (SM) of particle physics. The most spectacular finding would be the observation of resonant production of new particles that would show up as a bump in the invariant mass spectrum of certain observed final states.

The ATLAS and CMS Collaborations have recently reported a small excess over the expected diphoton mass spectrum, in the first  13~TeV collisions recorded at the LHC~\cite{CMS_note,ATLAS_note}.  The excess lies at an invariant mass of approximately $\sim 750$~GeV, 
with a decay width estimated to $\Gamma^{\rm tot}\leq 45 $~GeV by the experimental analyses~\cite{CMS_Moriond,ATLAS_Moriond}.
 While it is too early at this stage to know whether this excess is real or if it is due to statistical fluctuations,
it is important to discuss which particle beyond the SM  might explain the excess and how to test such hypotheses further. Many suggestions have been recently proposed, see Refs.~\cite{
Harigaya:2015ezk,
Nakai:2015ptz,
Mambrini:2015wyu,
Backovic:2015fnp,
Angelescu:2015uiz,
Knapen:2015dap,
Buttazzo:2015txu,
Pilaftsis:2015ycr,
Franceschini:2015kwy,
DiChiara:2015vdm,
Higaki:2015jag,
McDermott:2015sck,
Ellis:2015oso,
Low:2015qep,
Bellazzini:2015nxw,
Gupta:2015zzs,
Petersson:2015mkr,
Molinaro:2015cwg,
Bai:2015nbs,
Aloni:2015mxa,
Falkowski:2015swt,
Fichet:2015vvy,
Csaki:2015vek,
Chakrabortty:2015hff,
Bian:2015kjt,
Curtin:2015jcv,
Chao:2015ttq,
Demidov:2015zqn,
No:2015bsn,
Becirevic:2015fmu,
Martinez:2015kmn,
Agrawal:2015dbf,
Ahmed:2015uqt,
Cox:2015ckc,
Kobakhidze:2015ldh,
Matsuzaki:2015che,
Cao:2015pto,
Benbrik:2015fyz,
Kim:2015ron,
Gabrielli:2015dhk,
Alves:2015jgx,
Bernon:2015abk,
Dhuria:2015ufo,
Han:2015cty,
Han:2015dlp,
Luo:2015yio,
Chang:2015sdy,
Bardhan:2015hcr,
Feng:2015wil,
Barducci:2015gtd,
Chao:2015nsm,
Chakraborty:2015jvs,
Ding:2015rxx,
Han:2015qqj,
Wang:2015kuj,
Cao:2015twy,
Huang:2015evq,
Heckman:2015kqk,
Bi:2015uqd,
Kim:2015ksf,
Cline:2015msi,
Bauer:2015boy,
Chala:2015cev,
deBlas:2015hlv,
Boucenna:2015pav,
Murphy:2015kag,
Hernandez:2015ywg,
Dey:2015bur,
Pelaggi:2015knk,
Cao:2015xjz,
Huang:2015rkj,
Patel:2015ulo,
Chakraborty:2015gyj,
Altmannshofer:2015xfo,
Cvetic:2015vit,
Allanach:2015ixl,
Das:2015enc,
Cheung:2015cug,
Liu:2015yec,
Zhang:2015uuo,
An:2015cgp,
Wang:2015omi,
Cao:2015scs,
Gao:2015igz,
Goertz:2015nkp,
Dev:2015vjd,
Li:2015jwd,
Son:2015vfl,
Tang:2015eko,
Cao:2015apa,
Cai:2015hzc,
Chao:2015nac,
Anchordoqui:2015jxc,
Bizot:2015qqo,
Ibanez:2015uok,
Kang:2015roj,
Kanemura:2015bli,
Low:2015qho,
Kaneta:2015qpf,
Dasgupta:2015pbr,
Jung:2015etr,
Nomura:2016fzs,
Ko:2016lai,
Palti:2016kew,
Han:2016bus,
Danielsson:2016nyy,
Chao:2016mtn,
Csaki:2016raa,
Karozas:2016hcp,
Dutta:2016jqn,
Deppisch:2016scs,
Ito:2016zkz,
Kanemura:2015vcb, 
Bi:2015lcf, 
Antipin:2015kgh, 
Hamada:2015skp, 
Dev:2015isx, 
Hernandez:2016rbi, 
Ghorbani:2016jdq,  
Modak:2016ung}.

\begin{figure}
\includegraphics[width=\columnwidth]{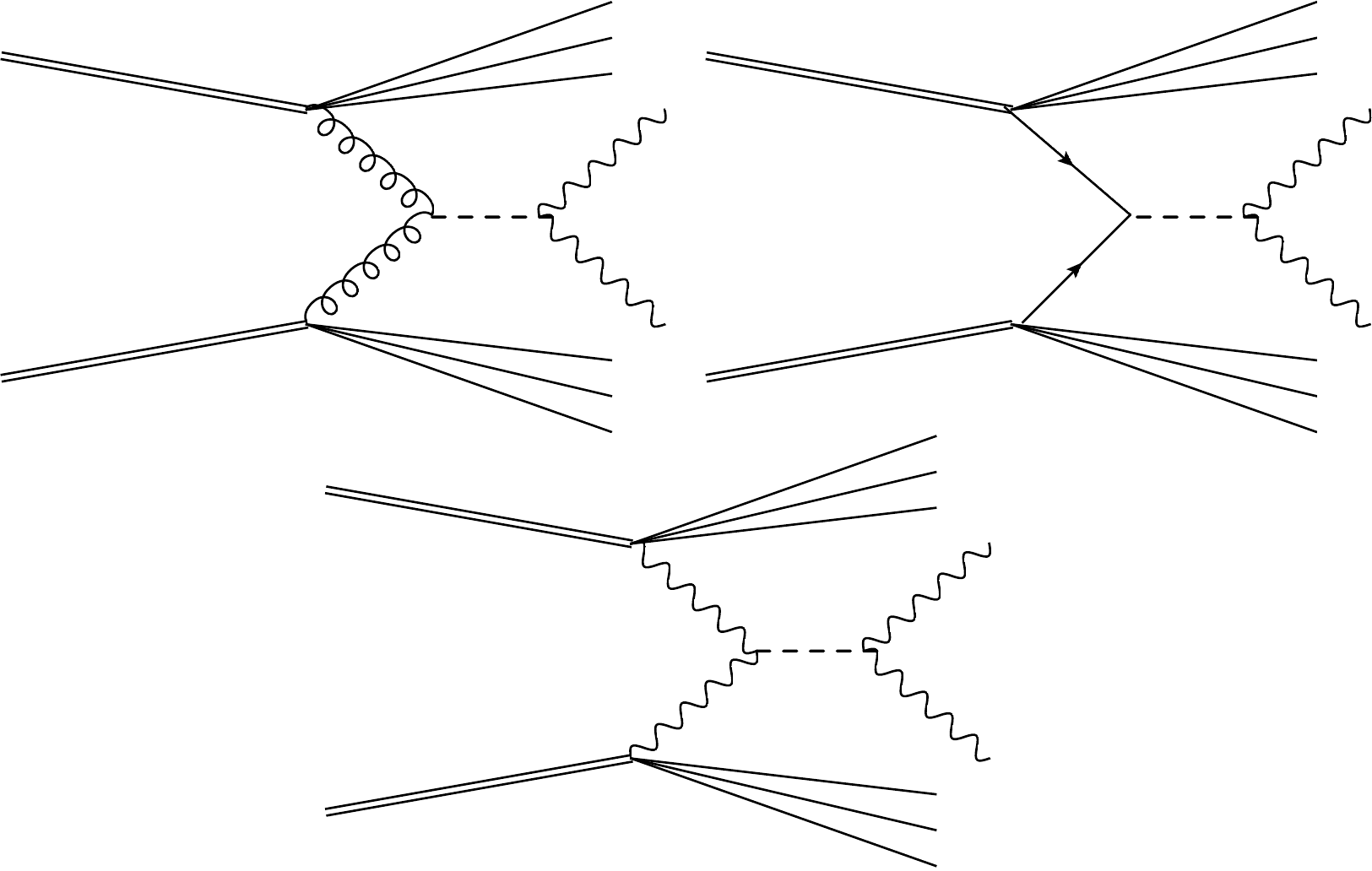}
\caption{\label{fig:inelastic}
Schematic representation of the resonant inelastic process $pp\to\gamma\gamma X$ with gluon and quark fusion (above) and photon fusion (below).}
\end{figure}

In this letter we will work under the assumption that the excess is due to a spin-0 resonance which we will denote by $\phi$. 
The next step is to pin down its properties, in particular how it couples to SM fields. 
One possibility is to investigate other potential decay channels, in particular decays into  $ZZ$, $Z\gamma$ and $W^+W^-$ are generically expected \cite{Fichet:2015vvy,Low:2015qho}. 
On the other hand, as with the SM Higgs boson, a lot of information could be obtained if one were able to tag individual production modes. 
Most of the recent literature has been focussing on gluon fusion or quark fusion (see,~e.g.~\cite{Franceschini:2015kwy}). Given that the resonance has to have sizable couplings to photons, another possibility is photon fusion 
\cite{Fichet:2015vvy,Csaki:2015vek,Anchordoqui:2015jxc,Csaki:2016raa}.
These production modes are dominantly inelastic, as the protons are destroyed in the collision, as depicted in Fig.~\ref{fig:inelastic}.

\begin{figure}
\includegraphics[width=\columnwidth]{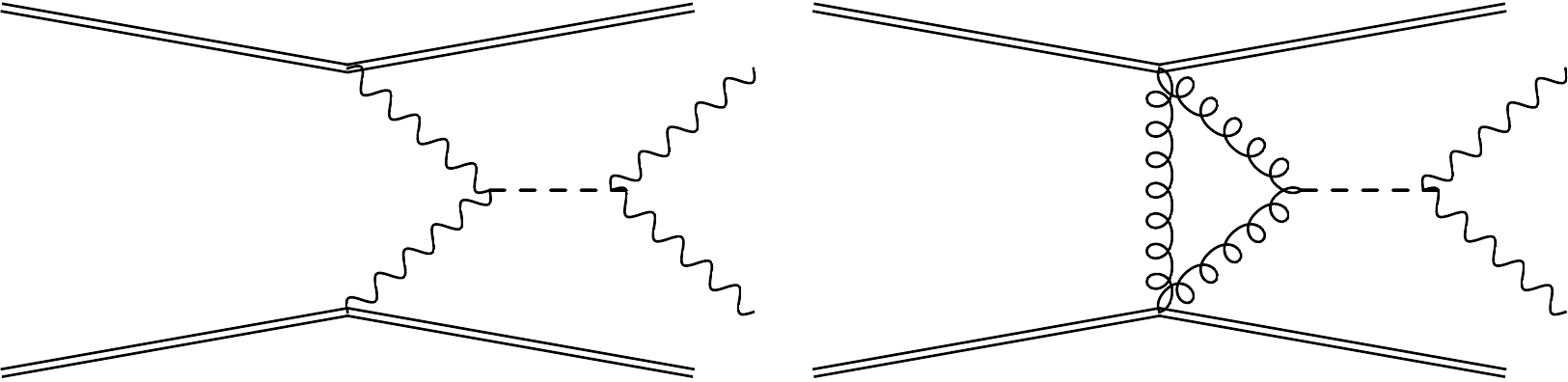}
\caption{\label{fig:elastic}
Schematic representation of the resonant elastic process $pp\to \gamma\gamma pp$. The elastic gluon fusion process  requires an additional exchange of a virtual gluon.
}
\end{figure}

In this letter we propose to measure directly the coupling of the resonance to photons in the elastic scattering process $pp\to pp\gamma\gamma$, in which the colliding protons remain intact.
For this we have to consider the two diagrams in Fig.~\ref{fig:elastic}. The first one is photon fusion, the second one is  gluon fusion with an additional gluon exchange to ensure that no color is extracted from the proton. 
As will be shown below, for any set of parameters explaining the diphoton excess at 750 GeV, the second process will have too small a cross section. In turn, using the former process, namely \textit{elastic photon fusion}, it will be possible to directly measure the photon coupling to the resonance with a precision that allows to access the theoretically interesting parameter space.
The experimental strategy to suppress the dominant inelastic processes and thereby allowing us to observe the elastic photon fusion process is to demand the detection of intact protons in forward detectors. 
In this way,  a very clean sample of exclusive di-photon production can be obtained, and requesting a good matching 
between the diphoton kinematical properties (mass and rapidity) as measured in the central CMS or ATLAS detectors and the intact protons measured in CT-PPS or AFP removes almost completely the background~\cite{Fichet:2014uka,Fichet:2015yba}.  

We would like to stress that in our proposal to measure the coupling of the resonance to photons  we do not make any a priori assumption about which of the three production modes in Fig.~\ref{fig:inelastic} is mainly responsible for the observed excess. The reason is that the forward tagging allows one to suppress all of these production modes equally, and one is just left with the first diagram of Fig.~\ref{fig:elastic}.

\section{Effective couplings and experimental constraints}

In this section we would like to give a brief overview of the possible production modes for the 750 GeV resonance and their  implications for the strength of the coupling to photons. For concreteness we consider two typical values for the total width $\Gamma^{\rm tot}=0.5$ GeV and $45$ GeV.

Let us parameterize the most general linear couplings of the 750 GeV resonance $\phi$ to the SM gauge fields and quarks by the effective Lagrangian of Ref.~\cite{Fichet:2015yia}
\bea
\mathcal L&=& \phi\left(\frac{1}{f_g}(G_{\mu\nu})^2+\frac{1}{f_B}(B_{\mu\nu})^2+\frac{1}{f_W}(W_{\mu\nu})^2+\frac{1}{f_H}|D_\mu H|^2
\right.\nn\\
&&
\left.
-\frac{1}{f_u}Y^u_{ij}H\bar q^i_L u^j_R
-\frac{1}{f_d}Y^d_{ij}H\bar q^i_L d^j_R+h.c.
\right)
\label{eft}
\eea 
where $G$, $W$ and $B$ denote the SM gauge fields, $H$ the Higgs, and $q^i$, $d^i$ and $u^i$ the quarks. The matrices $Y^{u,d}$ are the SM Yukawa couplings \footnote{For simplicity and to reduce the number of free parameters, we are assuming minimal flavour violation.
 This assumption is however not crucial for our present study.}.
The operator $\phi |D_\mu H|^2$ can generate couplings to longitudinal gauge bosons and the Higgs, but not to photons. It will be neglected in what follows, as its only effect for our purposes will be a contribution to the width of $\phi$.

After electroweak symmetry breaking, the coupling to photons $\mathcal L_{\phi\gamma\gamma}=f_\gamma^{-1}\phi\,(F_{\mu\nu})^2$ is given by 
\be
f_\gamma^{-1}=c^2_w\, f_{B}^{-1}+s^2_w\, f_W^{-1}\,.
\label{fgamma}
\ee
The expected strength of the coupling $f_\gamma^{-1}$ depends on the various production modes of $\phi$.
For $f_{g,u,d}^{-1}$ very small or zero, pure (inelastic) photon fusion dominates. In this case one can robustly translate the measured excess as \cite{Fichet:2015vvy}
\begin{eqnarray}
f_{\gamma}\approx 13.4 {\rm\ TeV} \quad (\Gamma^{\rm tot}=0.5~\textrm{GeV}) \nonumber \\
f_{\gamma}\approx 4.4 {\rm\ TeV} \quad (\Gamma^{\rm tot}=45~\textrm{GeV})
\label{photonfusion}
\end{eqnarray}
with $68\%$ credible region of $3.9 - 4.9$ TeV and $12.9 - 15.1$ TeV respectively.

\begin{figure}
\begin{picture}(200,330)
\put(0,0){\includegraphics[width=5.5 cm]{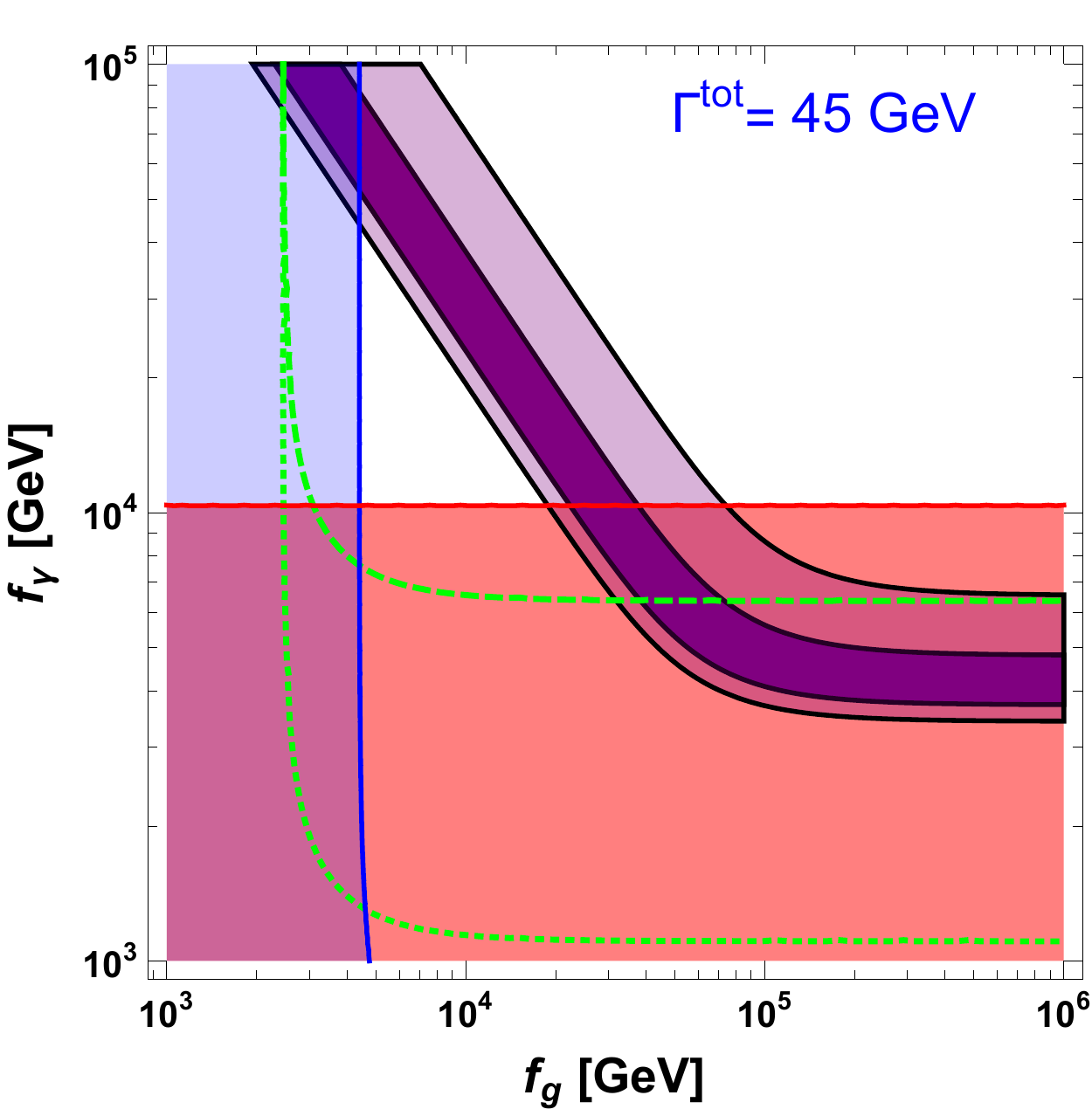}}
\put(0,180){\includegraphics[width=5.5 cm]{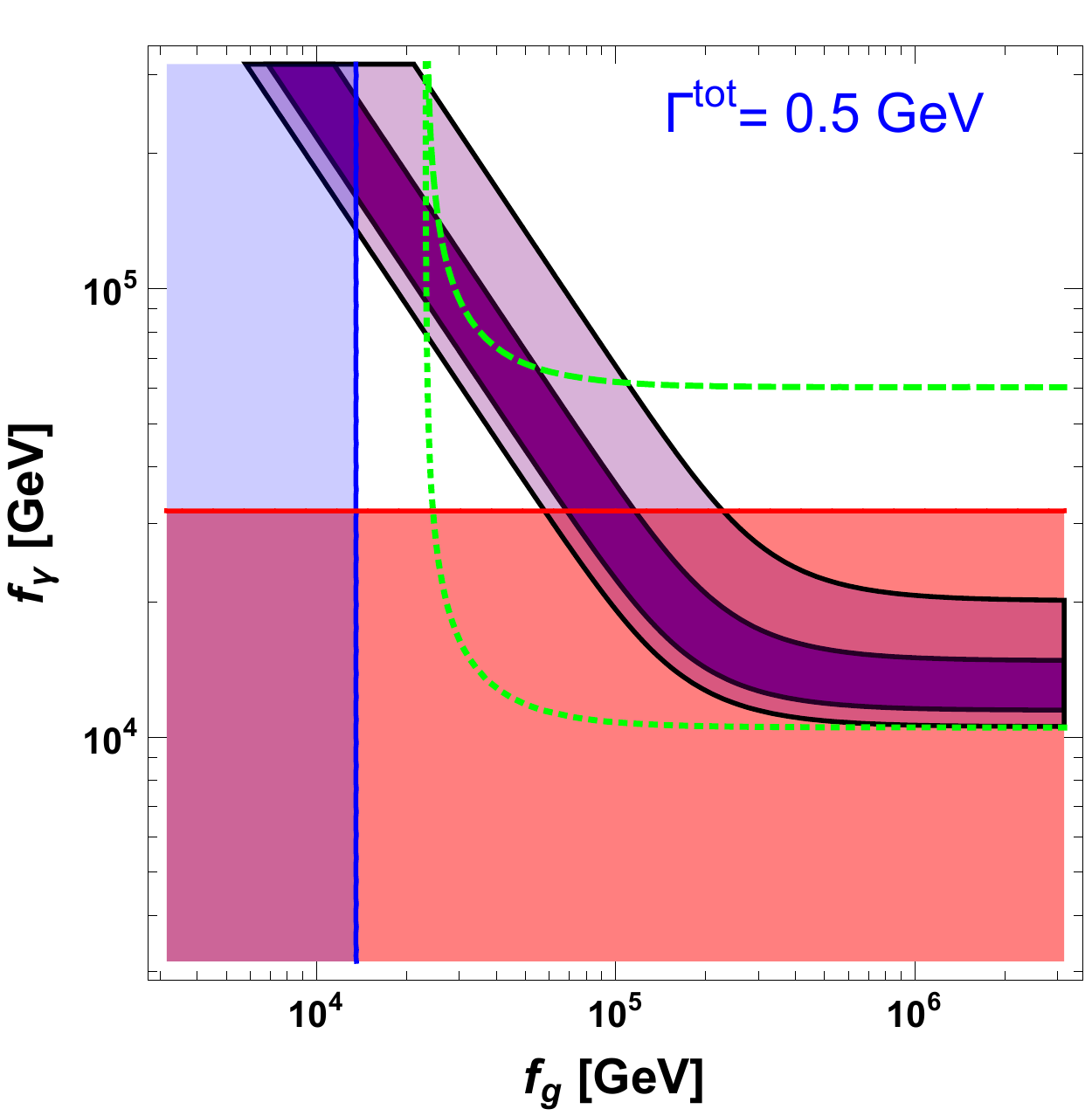}}
\end{picture}
\caption{
Bounds and sensitivities in the $f_\gamma-f_g$ plane, in case of production via photon and gluon fusion. 
\textit{Purple}:  $68\%$ C.L. and  $95\%$ C.L. credible regions corresponding to the observed diphoton event rate.
\textit{Green lines}: Limit of the region above which $\Gamma_{EW}+\Gamma_{gg} \leq\Gamma^{\rm tot}$.   Dotted (dashed) lines correspond to
$\Gamma_{EW}/\Gamma_{\gamma\gamma}=1.64$ (53.9) respectively. 
\textit{Blue}: Excluded region from Run 1 dijet searches \cite{ATLAS_dijet_8TeV,CMS_dijet_8TeV}. 
\textit{Red}: Sensitivity region from the potential measurement of $pp\rightarrow \gamma\gamma pp $ using forward proton detectors, for $300$~fb$^{-1}$ of integrated luminosity, see Eq.~(\ref{eq:sens}).
\label{fig:gluon_photon}
}
\end{figure}

Once the coupling $f_g^{-1}$ is increased, gluon fusion starts to dominate over photon fusion.
The allowed region in the plane $f_g$ - $f_\gamma$ is depicted in Fig.~\ref{fig:gluon_photon}.
The decay width into electroweak bosons and gluons $\Gamma_{\rm EW}+\Gamma_{gg}$ is required not to exceed the observed total width. The electroweak width is given by  $\Gamma_{\rm EW}=\Gamma_{\gamma\gamma}+\Gamma_{\gamma Z}+\Gamma_{ZZ}+\Gamma_{WW}$ and satisfies $1.64< \Gamma_{\rm EW}/\Gamma_{\gamma\gamma} <53.9$ from theoretical  and experimental constraints (see Ref.~\cite{Fichet:2015vvy} and Fig.~\ref{fig:XS_VV}), 
 
The weak coupling region ($f_\gamma$ large) requires large couplings to gluons to compensate the small branching fraction into photons. This region can be probed with dijet searches.
One can see that the expected strength of the photon coupling varies roughly between
\begin{eqnarray}
f_{\gamma}\approx 14\dots 248 {\rm\ TeV} \quad (\Gamma^{\rm tot}=0.5~\textrm{GeV}) \nonumber \\
f_{\gamma}\approx 4.7\dots 80{\rm\ TeV} \quad (\Gamma^{\rm tot}=45~\textrm{GeV})
\label{gluonfusion}
\end{eqnarray}
at $68\%$ CL.
Our method will be able to probe the strong coupling region ($f_\gamma$ small) that is insensitive to dijet searches.



\section{Experimental setup}

The strategy we propose to measure elastic diphoton production (see Fig.~\ref{fig:elastic}) relies on the observation of intact protons in the final state
using the AFP and CT-PPS forward proton detectors.
Simultaneously, the two photons are measured in the central CMS and 
ATLAS detectors.
 The forward detectors are located symmetrically at about 210~m from the main 
interaction point and cover the range $0.015 < \xi < 0.15$, 
where $\xi$ is the proton fractional  momentum loss, for the standard LHC lattice used at high luminosity. For a  $\sim 750$ GeV resonance produced  in  13 TeV collisions, one expects $\xi\sim 0.06 $.
At the LHC, a large number of interactions (called pile up) occurs within the same bunch crossing in order to obtain a large luminosity. 
Given the fact that the SM exclusive production cross section of two photons is very small \cite{Fichet:2014uka}, the main background originates from pile up, i.e. 
the production of two photons  superimposed with an elastic soft event producing two intact protons.  
 The proton tagging allows us to 
require a good matching between the proton and di-photon kinematical properties, which in turn reduces the pile up background to a small amount estimated to \be \sigma^{\rm bkd}_{\gamma\gamma p p}=3\cdot 10^{-4}~\textrm{fb} \label{eq:bg} \ee in \cite{Fichet:2014uka,Fichet:2015yba}.
In addition, the time-of-flight of the scattered proton could be measured with a precision of $\sim 10-15$~ps that provides a reconstruction of the interaction point of the protons within 2.1~mm inside ATLAS/CMS.  Checking
if the proton and photon scattering points are the same provides another way of suppressing the pile up background \footnote{We will not assume the use of timing detectors for the simulations in this paper}.

\section{Sensitivity to the diphoton coupling}

We will now estimate the sensitivity of the elastic scattering process to the diphoton coupling of the resonance.
We implement all gluon and photon initiated processes in the 
Forward Physics Monte Carlo (FPMC) Generator~\cite{FPMC}.  In case of the 
 two-photon $pp$ events, we use the 
 Budnev flux \cite{Budnev:1974de} which describes properly the coupling of the photon to the proton, taking into account the proton electromagnetic structure. 
 The survival probability of the colliding protons is expected to be close to 1 \cite{kmr}, here we implement a factor of $S^2\sim 0.72$ \cite{Harland-Lang:2016qjy}.
 The exclusive production via gluon exchanges is performed following the calculations by Khoze, Martin and Ryskin~\cite{kmr}. 
 The forward and central detector acceptance and resolution have been taken into account using a simplified simulation of the detector~\cite{Fichet:2014uka,
Fichet:2013gsa}, including realistic efficiencies for the central detector. The acceptance for the forward detectors is taken to be $0.015<\xi<0.15$, with 100\% efficiency in this window.

We will first argue that \textit{elastic gluon fusion} (EGGF), i.e. the second diagram in Fig.~\ref{fig:elastic}, can always be neglected for the excess under consideration. 
 The reason that this process is so small is due to the fact that the soft gluon emission in the gluon ladder has to be suppressed
in order to get an exclusive diffractive event with intact protons. Technically, a Sudakov form factor is introduced to suppress this emission that kills the 
 cross section at high mass.

For a more quantitative estimate of this effect, consider the production cross section of a $750$ GeV SM Higgs via the same mechanism,
\be
\sigma_{\rm EGGF}^h\approx 2\cdot 10^{-3} {\ \rm fb}\,.
\ee
Moreover, we also know the inelastic gluon fusion cross-section for a $750$ GeV SM Higgs  \cite{Dittmaier:2011ti},
\be
\sigma_{\rm GGF}^{h}\approx 550\ {\rm fb}\,.
\ee
We can now recast these cross-sections  in order to put a bound on the resonant production of $\phi$ by EGGF. 
Since the total GGF production cross section of the scalar resonance cannot exceed the observed cross section of the 750 GeV resonance, $\sigma_{\rm GGF}\,{\cal B}_{\gamma\gamma}<10$ fb, one can easily bound
\be
\sigma_{\rm EGGF}\,{\cal B}_{\gamma\gamma}=\frac{\sigma_{\rm EGGF}^{h}}{\sigma_{\rm GGF}^h}\sigma_{\rm GGF}\,{\cal B}_{\gamma\gamma}<3 \cdot 10^{-5} {\ \rm fb}.
\ee 
It follows that the elastic gluon fusion process is extremely small. Therefore, only the process of elastic photon fusion remains. This process provides a direct access to the photon coupling of the resonance,  i.e.~the quantity $f_\gamma$ in Eq.~(\ref{fgamma}).

We obtain a cross section of 
\be
\sigma_{pp\rightarrow \gamma\gamma pp}=  ( 0.184~{\rm fb})\, \left(\frac{5\, {\rm TeV}}{f_{\gamma}}\right)^4 \left(\frac{45~{\rm GeV}}{\Gamma_{\rm tot}}\right) \,.
\label{eq:XS2}
\ee
This cross section readily provides the sensitivity of the $\gamma\gamma pp$ measurement  to the diphoton coupling of the scalar resonance.

\begin{figure}
\includegraphics[width=8 cm]{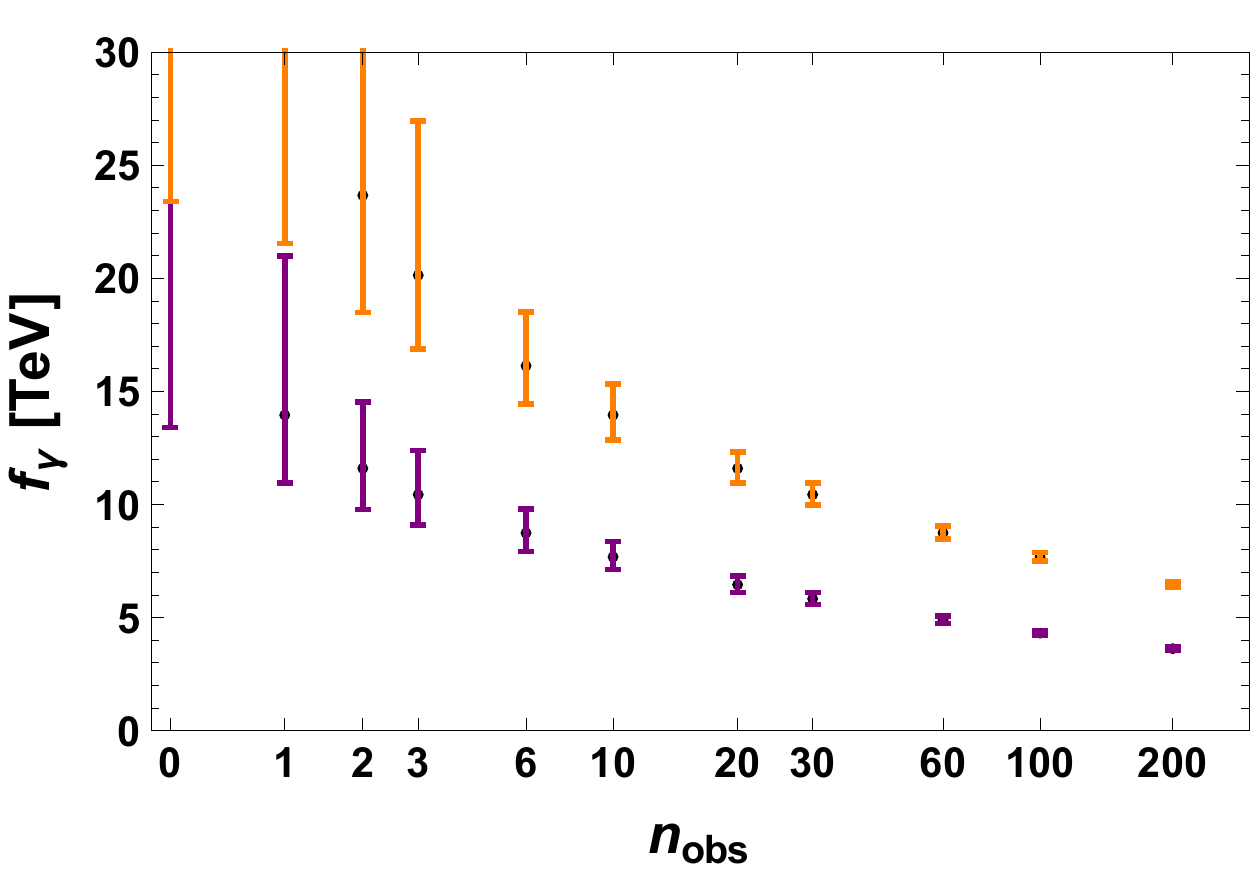}
\caption{
Inferred value of $f_\gamma$ (68\% C.L.) as a function of the observed number of events for 300 fb$^{-1}$ (purple) and 3000 fb$^{-1}$ (orange) of data. We have assumed $\Gamma^{\rm tot}=45$ GeV.
\label{fig:fgam_plot}
}
\end{figure}

Assuming Poisson statistics and the background rate Eq.~\eqref{eq:bg}, one can readily infer the values of $f_\gamma$ for given luminosity and number of observed events (see Fig.\ref{fig:fgam_plot}). One can also obtain the exclusion bound on $f_\gamma$ in the absence of any events, we find
\begin{eqnarray}
f_\gamma>41.4\ (31.9) {\rm \ TeV} \quad (\Gamma^{\rm tot}=0.5~\textrm{GeV}) \nonumber \\
f_\gamma>13.4\ (10.4) {\rm \ TeV} \quad (\Gamma^{\rm tot}=45~\textrm{GeV}) 
\label{eq:sens}
\end{eqnarray}
at 68\% (95\%) C.L. and 300 fb$^{-1}$. We use the 95\% C.L.~bound as a definition for the sensitivity of our method and also show it in Fig.~\ref{fig:gluon_photon}. For $3000$~fb$^{-1}$, the sensitivity would reach $f_\gamma>61.4 $ and $19.9\ {\rm \ TeV} $  respectively for the small and large width cases.  One observes that this elastic measurement is \textit{complementary} to dijet searches which typically probe the weak diphoton coupling regime.
With enough luminosity, both measurements together should give  access to  the whole relevant parameter space (purple regions).

 
\section{Prospects for elastic gauge boson production}

\begin{figure}
\includegraphics[width=8 cm]{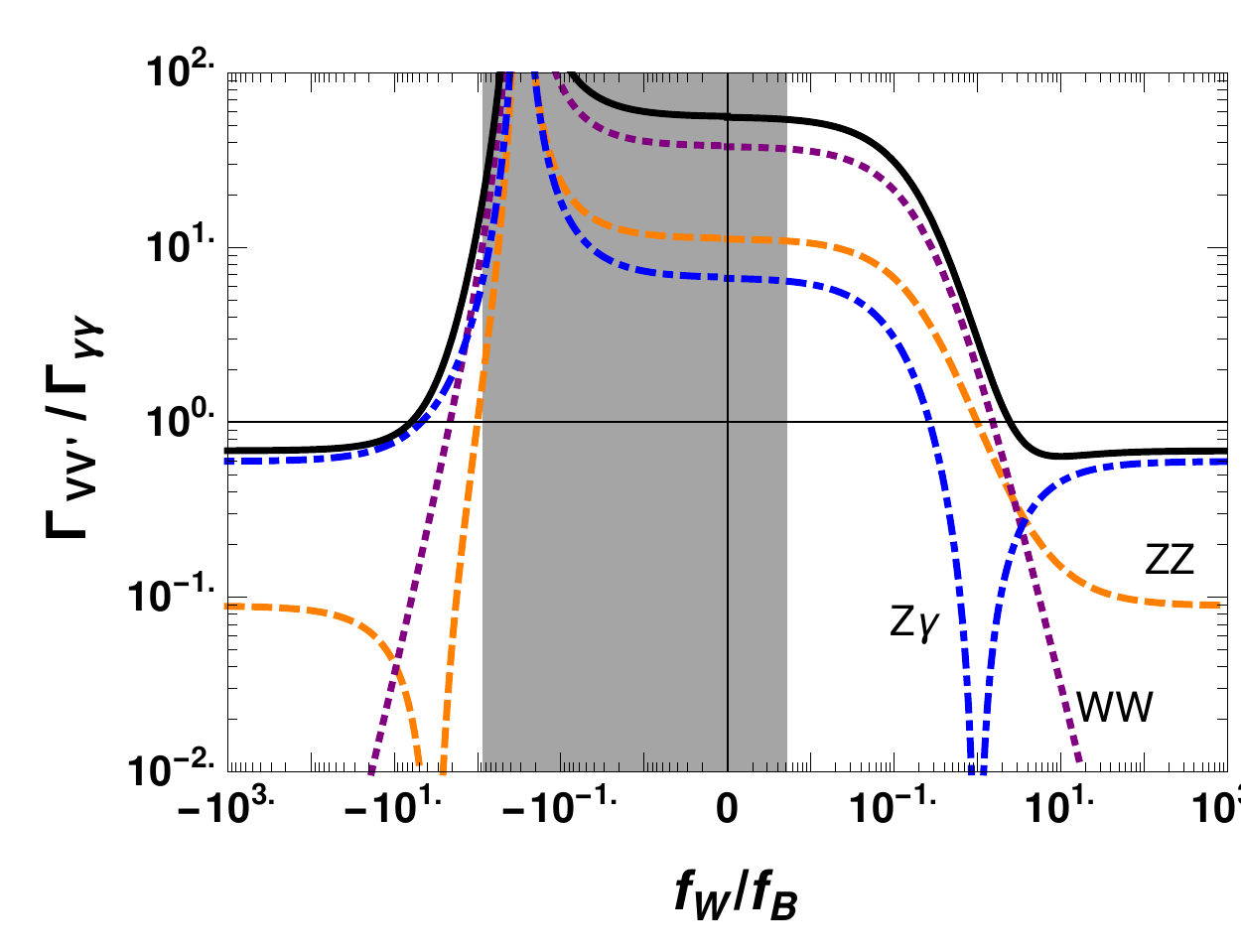}
\caption{Partial decay width for $\phi \to ZZ, Z\gamma, W^+W^-$ normalised to the $\phi \to \gamma\gamma$ width. The black line corresponds to $\Gamma_{\rm EW}/\Gamma_{\gamma\gamma}$.
The grey region is excluded at $95\%$ C.L. by the $Z\gamma$ search from Run 1 \cite{Aad:2014fha}.
\label{fig:XS_VV}
}
\end{figure}

It is clear from $SU(2)\times U(1)_Y$ gauge invariance that the diphoton coupling of the resonance must be accompanied by a coupling to $ZZ$ , to $Z\gamma$ and potentially $W^+W^-$. From the effective couplings of Eq.~(\ref{eft}), there are two independent operators $f_B^{-1}\phi (B_{\mu\nu})^2$, $f_W^{-1}\phi(W_{\mu\nu})^2$. The partial decay widths into weak bosons can be found in \cite{Fichet:2015vvy} and are shown in Fig.~\ref{fig:XS_VV}. The ratio $\Gamma_{\rm EW}/\Gamma_{\gamma\gamma}$ is bounded from above from diboson searches at LHC Run 1 \cite{Aad:2014fha,Aad:2015kna,Khachatryan:2015cwa,Aad:2015agg}. We show an exclusion bound in Fig.~\ref{fig:XS_VV}, obtained by taking the lowest $95\%$ C.L. value  $\sigma_{pp\rightarrow \gamma\gamma X}^{13{\rm TeV}}=2.5 $~fb and assuming a typical factor $\sim 4$ with respect to the $8$ TeV rate. The $Z\gamma $ bound \cite{Aad:2014fha} from Run 1 turns out to be the most stringent one, excluding the $-0.87<f_W/f_B<0.005$ region, implying in particular $\Gamma_{EW}/\Gamma_{\gamma\gamma}<53.9$.


The ratios of event rates $\sigma_{pp\to VV' X}/\sigma_{pp\to \gamma\gamma X}$ in case of  production via gluon fusion and quark fusion are proportional to $\Gamma_{VV'}/\Gamma_{\gamma\gamma}$. This is also the case for elastic production, the $W$ and $Z$ fluxes  from the proton being negligible 
\footnote{This can be also true for inelastic electroweak boson fusion, to the extent that the $WW$, $ZZ$, $Z\gamma$ fusion diagrams are dominated by inelastic photon fusion, which typically occurs, and provided that the contribution of the $\phi |D^\mu H|$ operator is negligible.}.
Whenever this condition  is true,  the measurement of one of the $ZZ$, $Z\gamma$, $WW$ rates readily provides access to the two couplings $f_B^{-1}$, $f_W^{-1}$.
Interestingly, the event rates into other gauge boson pairs can be substantially larger than the photon-photon one, in particular if the coupling to the $(W_{\mu\nu})^2$ operator dominates.

Searches in inelastic channels would be one evident method to pin down the $f_B^{-1}$, $f_W^{-1}$ couplings. However, just like for $\gamma\gamma$, the elastic $VV'$ channels are also of interest because  they contain information which is complementary from the inelastic ones.
Let us briefly comment  about such elastic searches.
\begin{itemize}
\item $pp\rightarrow ZZpp$:   At least one  Z decaying leptonically has to be required because of the huge QCD background. The other $Z$ can be tagged as a large-radius jet. However, because of the small branching ratio ($\sim 9\%$), and after taking account selection efficiencies, this channel would hardly be competitive with the diphoton one. 

\item $pp\rightarrow Z\gamma pp$:  The large-radius jet arising from the hadronic decay of the  $Z$ can be efficiently tagged using increasingly powerful jet substructure techniques. Using the full kinematic information provided by forward proton detection, i.e.~matching the jet - photon system with the proton-proton system, an excellent background rejection is expected. It is expected to be slightly lower than in the diphoton case because of the worse resolution on the jet momentum. As the event rates can be up to $6.4$ times larger than the $\gamma\gamma$ case, this channel is potentially competitive with the diphoton channel. A full study including all pile up background is worth to be considered.

\item $pp\rightarrow W^+W^- pp$:  Requesting a fully leptonic decay of the $WW$ pair implies an overall branching ratio of $\sim 10\%$. This is potentially interesting as the $WW$ rate can be up to $\sim 37$ times larger than the $\gamma\gamma$ rate. There is a background at the matrix element level, the main one being SM dilepton production via $\gamma\gamma\rightarrow \ell\ell$. This background can be completely removed by asking non back-to-back leptons, using a cut on the azimuthal angle between leptons 
\cite{Chapon:2009hh,Kepka:2008yx}. 
Removing the pile up background requires the installation of precise timing detectors 
since the matching between the $WW$ and proton informations lacks efficiency because of the presence of the two neutrinos. Although a detailed study is needed to evaluate the potential of this channel, one may expect that $WW$ searches are potentially  competitive with respect to the $\gamma\gamma$ searches.

\end{itemize} 


\section{Conclusion}

We have demonstrated that the diphoton coupling $f_\gamma^{-1}\phi\,(F_{\mu\nu})^2$ of a putative $750$ GeV resonance $\phi$ can be accurately determined in a completely model-independent way by tagging the elastic process $pp\to\gamma\gamma pp$ 
with forward detectors, thereby obtaining a background-free sample of photon-induced processes. We find a sensitivity $f_\gamma\approx$ 31.5 TeV (10.2 TeV) at 300 fb$^{-1}$ for $\Gamma^{\rm tot}=0.5 (45)$ GeV, covering a large portion of the parameter space of models predicting a production of $\phi$ in gluon or photon fusion. Notice that our method alone cannot exclude models with $f_g\lesssim f_\gamma$.

Provided  that the total width is independently measured, the determination of $f_\gamma$ provides indirect information about the dominant production mode. For instance, if $\Gamma^{\rm tot}\approx  45$ GeV, $f_\gamma>5$ TeV would exclude  photon fusion as the main production mechanism. Further techniques of how to disentangle the production mechanisms have recently been discussed in Ref.~\cite{Csaki:2016raa}.

Furthermore we have commented on the various other channels that one can probe with elastic measurements. Detecting the $Z\gamma$ final state  seems possible, as well as the $WW$ final state provided that timing detectors can be exploited.

\begin{acknowledgments}
S.F. acknowledge the Funda\c c\~ao de Amparo \`a Pesquisa do Estado de S\~ao Paulo (FAPESP) for financial support. 

\end{acknowledgments}

\bibliography{paper,diphotonpapers}

\providecommand{\noopsort}[1]{}\providecommand{\singleletter}[1]{#1}%
\begin{thebibliography}{144}%
\makeatletter
\providecommand \@ifxundefined [1]{%
 \@ifx{#1\undefined}
}%
\providecommand \@ifnum [1]{%
 \ifnum #1\expandafter \@firstoftwo
 \else \expandafter \@secondoftwo
 \fi
}%
\providecommand \@ifx [1]{%
 \ifx #1\expandafter \@firstoftwo
 \else \expandafter \@secondoftwo
 \fi
}%
\providecommand \natexlab [1]{#1}%
\providecommand \enquote  [1]{``#1''}%
\providecommand \bibnamefont  [1]{#1}%
\providecommand \bibfnamefont [1]{#1}%
\providecommand \citenamefont [1]{#1}%
\providecommand \href@noop [0]{\@secondoftwo}%
\providecommand \href [0]{\begingroup \@sanitize@url \@href}%
\providecommand \@href[1]{\@@startlink{#1}\@@href}%
\providecommand \@@href[1]{\endgroup#1\@@endlink}%
\providecommand \@sanitize@url [0]{\catcode `\\12\catcode `\$12\catcode
  `\&12\catcode `\#12\catcode `\^12\catcode `\_12\catcode `\%12\relax}%
\providecommand \@@startlink[1]{}%
\providecommand \@@endlink[0]{}%
\providecommand \url  [0]{\begingroup\@sanitize@url \@url }%
\providecommand \@url [1]{\endgroup\@href {#1}{\urlprefix }}%
\providecommand \urlprefix  [0]{URL }%
\providecommand \Eprint [0]{\href }%
\providecommand \doibase [0]{http://dx.doi.org/}%
\providecommand \selectlanguage [0]{\@gobble}%
\providecommand \bibinfo  [0]{\@secondoftwo}%
\providecommand \bibfield  [0]{\@secondoftwo}%
\providecommand \translation [1]{[#1]}%
\providecommand \BibitemOpen [0]{}%
\providecommand \bibitemStop [0]{}%
\providecommand \bibitemNoStop [0]{.\EOS\space}%
\providecommand \EOS [0]{\spacefactor3000\relax}%
\providecommand \BibitemShut  [1]{\csname bibitem#1\endcsname}%
\let\auto@bib@innerbib\@empty
\bibitem [{CMS(2015{\natexlab{a}})}]{CMS_note}%
  \BibitemOpen
  \href {https://cds.cern.ch/record/2114808} {\emph {\bibinfo {title} {{Search
  for new physics in high mass diphoton events in proton-proton collisions at
  13TeV}}}},\ \bibinfo {type} {Tech. Rep.}\ \bibinfo {number}
  {CMS-PAS-EXO-15-004}\ (\bibinfo  {institution} {CERN},\ \bibinfo {address}
  {Geneva},\ \bibinfo {year} {2015})\BibitemShut {NoStop}%
\bibitem [{ATL(2015)}]{ATLAS_note}%
  \BibitemOpen
  \href {http://cds.cern.ch/record/2114853} {\emph {\bibinfo {title} {{Search
  for resonances decaying to photon pairs in 3.2 fb$^{-1}$ of $pp$ collisions
  at $\sqrt{s}$ = 13 TeV with the ATLAS detector}}}},\ \bibinfo {type} {Tech.
  Rep.}\ \bibinfo {number} {ATLAS-CONF-2015-081}\ (\bibinfo  {institution}
  {CERN},\ \bibinfo {address} {Geneva},\ \bibinfo {year} {2015})\BibitemShut
  {NoStop}%
\bibitem [{CMS()}]{CMS_Moriond}%
  \BibitemOpen
  \href@noop {} {}\bibinfo {note} {{CMS diphoton presentation, Moriond
  Electroweak 2016}}\BibitemShut {NoStop}%
\bibitem [{ATL()}]{ATLAS_Moriond}%
  \BibitemOpen
  \href@noop {} {}\bibinfo {note} {{ATLAS diphoton presentation, Moriond
  Electroweak 2016}}\BibitemShut {NoStop}%
\bibitem [{\citenamefont {Harigaya}\ and\ \citenamefont
  {Nomura}(2015)}]{Harigaya:2015ezk}%
  \BibitemOpen
  \bibfield  {author} {\bibinfo {author} {\bibfnamefont {K.}~\bibnamefont
  {Harigaya}}\ and\ \bibinfo {author} {\bibfnamefont {Y.}~\bibnamefont
  {Nomura}},\ }\href@noop {} {\  (\bibinfo {year} {2015})},\ \Eprint
  {http://arxiv.org/abs/1512.04850} {arXiv:1512.04850 [hep-ph]} \BibitemShut
  {NoStop}%
\bibitem [{\citenamefont {Nakai}\ \emph {et~al.}(2015)\citenamefont {Nakai},
  \citenamefont {Sato},\ and\ \citenamefont {Tobioka}}]{Nakai:2015ptz}%
  \BibitemOpen
  \bibfield  {author} {\bibinfo {author} {\bibfnamefont {Y.}~\bibnamefont
  {Nakai}}, \bibinfo {author} {\bibfnamefont {R.}~\bibnamefont {Sato}}, \ and\
  \bibinfo {author} {\bibfnamefont {K.}~\bibnamefont {Tobioka}},\ }\href@noop
  {} {\  (\bibinfo {year} {2015})},\ \Eprint {http://arxiv.org/abs/1512.04924}
  {arXiv:1512.04924 [hep-ph]} \BibitemShut {NoStop}%
\bibitem [{\citenamefont {Mambrini}\ \emph {et~al.}(2015)\citenamefont
  {Mambrini}, \citenamefont {Arcadi},\ and\ \citenamefont
  {Djouadi}}]{Mambrini:2015wyu}%
  \BibitemOpen
  \bibfield  {author} {\bibinfo {author} {\bibfnamefont {Y.}~\bibnamefont
  {Mambrini}}, \bibinfo {author} {\bibfnamefont {G.}~\bibnamefont {Arcadi}}, \
  and\ \bibinfo {author} {\bibfnamefont {A.}~\bibnamefont {Djouadi}},\
  }\href@noop {} {\  (\bibinfo {year} {2015})},\ \Eprint
  {http://arxiv.org/abs/1512.04913} {arXiv:1512.04913 [hep-ph]} \BibitemShut
  {NoStop}%
\bibitem [{\citenamefont {Backovic}\ \emph {et~al.}(2015)\citenamefont
  {Backovic}, \citenamefont {Mariotti},\ and\ \citenamefont
  {Redigolo}}]{Backovic:2015fnp}%
  \BibitemOpen
  \bibfield  {author} {\bibinfo {author} {\bibfnamefont {M.}~\bibnamefont
  {Backovic}}, \bibinfo {author} {\bibfnamefont {A.}~\bibnamefont {Mariotti}},
  \ and\ \bibinfo {author} {\bibfnamefont {D.}~\bibnamefont {Redigolo}},\
  }\href@noop {} {\  (\bibinfo {year} {2015})},\ \Eprint
  {http://arxiv.org/abs/1512.04917} {arXiv:1512.04917 [hep-ph]} \BibitemShut
  {NoStop}%
\bibitem [{\citenamefont {Angelescu}\ \emph {et~al.}(2015)\citenamefont
  {Angelescu}, \citenamefont {Djouadi},\ and\ \citenamefont
  {Moreau}}]{Angelescu:2015uiz}%
  \BibitemOpen
  \bibfield  {author} {\bibinfo {author} {\bibfnamefont {A.}~\bibnamefont
  {Angelescu}}, \bibinfo {author} {\bibfnamefont {A.}~\bibnamefont {Djouadi}},
  \ and\ \bibinfo {author} {\bibfnamefont {G.}~\bibnamefont {Moreau}},\
  }\href@noop {} {\  (\bibinfo {year} {2015})},\ \Eprint
  {http://arxiv.org/abs/1512.04921} {arXiv:1512.04921 [hep-ph]} \BibitemShut
  {NoStop}%
\bibitem [{\citenamefont {Knapen}\ \emph {et~al.}(2015)\citenamefont {Knapen},
  \citenamefont {Melia}, \citenamefont {Papucci},\ and\ \citenamefont
  {Zurek}}]{Knapen:2015dap}%
  \BibitemOpen
  \bibfield  {author} {\bibinfo {author} {\bibfnamefont {S.}~\bibnamefont
  {Knapen}}, \bibinfo {author} {\bibfnamefont {T.}~\bibnamefont {Melia}},
  \bibinfo {author} {\bibfnamefont {M.}~\bibnamefont {Papucci}}, \ and\
  \bibinfo {author} {\bibfnamefont {K.}~\bibnamefont {Zurek}},\ }\href@noop {}
  {\  (\bibinfo {year} {2015})},\ \Eprint {http://arxiv.org/abs/1512.04928}
  {arXiv:1512.04928 [hep-ph]} \BibitemShut {NoStop}%
\bibitem [{\citenamefont {Buttazzo}\ \emph {et~al.}(2015)\citenamefont
  {Buttazzo}, \citenamefont {Greljo},\ and\ \citenamefont
  {Marzocca}}]{Buttazzo:2015txu}%
  \BibitemOpen
  \bibfield  {author} {\bibinfo {author} {\bibfnamefont {D.}~\bibnamefont
  {Buttazzo}}, \bibinfo {author} {\bibfnamefont {A.}~\bibnamefont {Greljo}}, \
  and\ \bibinfo {author} {\bibfnamefont {D.}~\bibnamefont {Marzocca}},\
  }\href@noop {} {\  (\bibinfo {year} {2015})},\ \Eprint
  {http://arxiv.org/abs/1512.04929} {arXiv:1512.04929 [hep-ph]} \BibitemShut
  {NoStop}%
\bibitem [{\citenamefont {Pilaftsis}(2015)}]{Pilaftsis:2015ycr}%
  \BibitemOpen
  \bibfield  {author} {\bibinfo {author} {\bibfnamefont {A.}~\bibnamefont
  {Pilaftsis}},\ }\href@noop {} {\  (\bibinfo {year} {2015})},\ \Eprint
  {http://arxiv.org/abs/1512.04931} {arXiv:1512.04931 [hep-ph]} \BibitemShut
  {NoStop}%
\bibitem [{\citenamefont {Franceschini}\ \emph {et~al.}(2015)\citenamefont
  {Franceschini}, \citenamefont {Giudice}, \citenamefont {Kamenik},
  \citenamefont {McCullough}, \citenamefont {Pomarol}, \citenamefont
  {Rattazzi}, \citenamefont {Redi}, \citenamefont {Riva}, \citenamefont
  {Strumia},\ and\ \citenamefont {Torre}}]{Franceschini:2015kwy}%
  \BibitemOpen
  \bibfield  {author} {\bibinfo {author} {\bibfnamefont {R.}~\bibnamefont
  {Franceschini}}, \bibinfo {author} {\bibfnamefont {G.~F.}\ \bibnamefont
  {Giudice}}, \bibinfo {author} {\bibfnamefont {J.~F.}\ \bibnamefont
  {Kamenik}}, \bibinfo {author} {\bibfnamefont {M.}~\bibnamefont {McCullough}},
  \bibinfo {author} {\bibfnamefont {A.}~\bibnamefont {Pomarol}}, \bibinfo
  {author} {\bibfnamefont {R.}~\bibnamefont {Rattazzi}}, \bibinfo {author}
  {\bibfnamefont {M.}~\bibnamefont {Redi}}, \bibinfo {author} {\bibfnamefont
  {F.}~\bibnamefont {Riva}}, \bibinfo {author} {\bibfnamefont {A.}~\bibnamefont
  {Strumia}}, \ and\ \bibinfo {author} {\bibfnamefont {R.}~\bibnamefont
  {Torre}},\ }\href@noop {} {\  (\bibinfo {year} {2015})},\ \Eprint
  {http://arxiv.org/abs/1512.04933} {arXiv:1512.04933 [hep-ph]} \BibitemShut
  {NoStop}%
\bibitem [{\citenamefont {Di~Chiara}\ \emph {et~al.}(2015)\citenamefont
  {Di~Chiara}, \citenamefont {Marzola},\ and\ \citenamefont
  {Raidal}}]{DiChiara:2015vdm}%
  \BibitemOpen
  \bibfield  {author} {\bibinfo {author} {\bibfnamefont {S.}~\bibnamefont
  {Di~Chiara}}, \bibinfo {author} {\bibfnamefont {L.}~\bibnamefont {Marzola}},
  \ and\ \bibinfo {author} {\bibfnamefont {M.}~\bibnamefont {Raidal}},\
  }\href@noop {} {\  (\bibinfo {year} {2015})},\ \Eprint
  {http://arxiv.org/abs/1512.04939} {arXiv:1512.04939 [hep-ph]} \BibitemShut
  {NoStop}%
\bibitem [{\citenamefont {Higaki}\ \emph {et~al.}(2015)\citenamefont {Higaki},
  \citenamefont {Jeong}, \citenamefont {Kitajima},\ and\ \citenamefont
  {Takahashi}}]{Higaki:2015jag}%
  \BibitemOpen
  \bibfield  {author} {\bibinfo {author} {\bibfnamefont {T.}~\bibnamefont
  {Higaki}}, \bibinfo {author} {\bibfnamefont {K.~S.}\ \bibnamefont {Jeong}},
  \bibinfo {author} {\bibfnamefont {N.}~\bibnamefont {Kitajima}}, \ and\
  \bibinfo {author} {\bibfnamefont {F.}~\bibnamefont {Takahashi}},\ }\href@noop
  {} {\  (\bibinfo {year} {2015})},\ \Eprint {http://arxiv.org/abs/1512.05295}
  {arXiv:1512.05295 [hep-ph]} \BibitemShut {NoStop}%
\bibitem [{\citenamefont {McDermott}\ \emph {et~al.}(2015)\citenamefont
  {McDermott}, \citenamefont {Meade},\ and\ \citenamefont
  {Ramani}}]{McDermott:2015sck}%
  \BibitemOpen
  \bibfield  {author} {\bibinfo {author} {\bibfnamefont {S.~D.}\ \bibnamefont
  {McDermott}}, \bibinfo {author} {\bibfnamefont {P.}~\bibnamefont {Meade}}, \
  and\ \bibinfo {author} {\bibfnamefont {H.}~\bibnamefont {Ramani}},\
  }\href@noop {} {\  (\bibinfo {year} {2015})},\ \Eprint
  {http://arxiv.org/abs/1512.05326} {arXiv:1512.05326 [hep-ph]} \BibitemShut
  {NoStop}%
\bibitem [{\citenamefont {Ellis}\ \emph {et~al.}(2015)\citenamefont {Ellis},
  \citenamefont {Ellis}, \citenamefont {Quevillon}, \citenamefont {Sanz},\ and\
  \citenamefont {You}}]{Ellis:2015oso}%
  \BibitemOpen
  \bibfield  {author} {\bibinfo {author} {\bibfnamefont {J.}~\bibnamefont
  {Ellis}}, \bibinfo {author} {\bibfnamefont {S.~A.~R.}\ \bibnamefont {Ellis}},
  \bibinfo {author} {\bibfnamefont {J.}~\bibnamefont {Quevillon}}, \bibinfo
  {author} {\bibfnamefont {V.}~\bibnamefont {Sanz}}, \ and\ \bibinfo {author}
  {\bibfnamefont {T.}~\bibnamefont {You}},\ }\href@noop {} {\  (\bibinfo {year}
  {2015})},\ \Eprint {http://arxiv.org/abs/1512.05327} {arXiv:1512.05327
  [hep-ph]} \BibitemShut {NoStop}%
\bibitem [{\citenamefont {Low}\ \emph {et~al.}(2015)\citenamefont {Low},
  \citenamefont {Tesi},\ and\ \citenamefont {Wang}}]{Low:2015qep}%
  \BibitemOpen
  \bibfield  {author} {\bibinfo {author} {\bibfnamefont {M.}~\bibnamefont
  {Low}}, \bibinfo {author} {\bibfnamefont {A.}~\bibnamefont {Tesi}}, \ and\
  \bibinfo {author} {\bibfnamefont {L.-T.}\ \bibnamefont {Wang}},\ }\href@noop
  {} {\  (\bibinfo {year} {2015})},\ \Eprint {http://arxiv.org/abs/1512.05328}
  {arXiv:1512.05328 [hep-ph]} \BibitemShut {NoStop}%
\bibitem [{\citenamefont {Bellazzini}\ \emph {et~al.}(2015)\citenamefont
  {Bellazzini}, \citenamefont {Franceschini}, \citenamefont {Sala},\ and\
  \citenamefont {Serra}}]{Bellazzini:2015nxw}%
  \BibitemOpen
  \bibfield  {author} {\bibinfo {author} {\bibfnamefont {B.}~\bibnamefont
  {Bellazzini}}, \bibinfo {author} {\bibfnamefont {R.}~\bibnamefont
  {Franceschini}}, \bibinfo {author} {\bibfnamefont {F.}~\bibnamefont {Sala}},
  \ and\ \bibinfo {author} {\bibfnamefont {J.}~\bibnamefont {Serra}},\
  }\href@noop {} {\  (\bibinfo {year} {2015})},\ \Eprint
  {http://arxiv.org/abs/1512.05330} {arXiv:1512.05330 [hep-ph]} \BibitemShut
  {NoStop}%
\bibitem [{\citenamefont {Gupta}\ \emph {et~al.}(2015)\citenamefont {Gupta},
  \citenamefont {JŠger}, \citenamefont {Kats}, \citenamefont {Perez},\ and\
  \citenamefont {Stamou}}]{Gupta:2015zzs}%
  \BibitemOpen
  \bibfield  {author} {\bibinfo {author} {\bibfnamefont {R.~S.}\ \bibnamefont
  {Gupta}}, \bibinfo {author} {\bibfnamefont {S.}~\bibnamefont {JŠger}},
  \bibinfo {author} {\bibfnamefont {Y.}~\bibnamefont {Kats}}, \bibinfo {author}
  {\bibfnamefont {G.}~\bibnamefont {Perez}}, \ and\ \bibinfo {author}
  {\bibfnamefont {E.}~\bibnamefont {Stamou}},\ }\href@noop {} {\  (\bibinfo
  {year} {2015})},\ \Eprint {http://arxiv.org/abs/1512.05332} {arXiv:1512.05332
  [hep-ph]} \BibitemShut {NoStop}%
\bibitem [{\citenamefont {Petersson}\ and\ \citenamefont
  {Torre}(2015)}]{Petersson:2015mkr}%
  \BibitemOpen
  \bibfield  {author} {\bibinfo {author} {\bibfnamefont {C.}~\bibnamefont
  {Petersson}}\ and\ \bibinfo {author} {\bibfnamefont {R.}~\bibnamefont
  {Torre}},\ }\href@noop {} {\  (\bibinfo {year} {2015})},\ \Eprint
  {http://arxiv.org/abs/1512.05333} {arXiv:1512.05333 [hep-ph]} \BibitemShut
  {NoStop}%
\bibitem [{\citenamefont {Molinaro}\ \emph {et~al.}(2015)\citenamefont
  {Molinaro}, \citenamefont {Sannino},\ and\ \citenamefont
  {Vignaroli}}]{Molinaro:2015cwg}%
  \BibitemOpen
  \bibfield  {author} {\bibinfo {author} {\bibfnamefont {E.}~\bibnamefont
  {Molinaro}}, \bibinfo {author} {\bibfnamefont {F.}~\bibnamefont {Sannino}}, \
  and\ \bibinfo {author} {\bibfnamefont {N.}~\bibnamefont {Vignaroli}},\
  }\href@noop {} {\  (\bibinfo {year} {2015})},\ \Eprint
  {http://arxiv.org/abs/1512.05334} {arXiv:1512.05334 [hep-ph]} \BibitemShut
  {NoStop}%
\bibitem [{\citenamefont {Bai}\ \emph {et~al.}(2015)\citenamefont {Bai},
  \citenamefont {Berger},\ and\ \citenamefont {Lu}}]{Bai:2015nbs}%
  \BibitemOpen
  \bibfield  {author} {\bibinfo {author} {\bibfnamefont {Y.}~\bibnamefont
  {Bai}}, \bibinfo {author} {\bibfnamefont {J.}~\bibnamefont {Berger}}, \ and\
  \bibinfo {author} {\bibfnamefont {R.}~\bibnamefont {Lu}},\ }\href@noop {} {\
  (\bibinfo {year} {2015})},\ \Eprint {http://arxiv.org/abs/1512.05779}
  {arXiv:1512.05779 [hep-ph]} \BibitemShut {NoStop}%
\bibitem [{\citenamefont {Aloni}\ \emph {et~al.}(2015)\citenamefont {Aloni},
  \citenamefont {Blum}, \citenamefont {Dery}, \citenamefont {Efrati},\ and\
  \citenamefont {Nir}}]{Aloni:2015mxa}%
  \BibitemOpen
  \bibfield  {author} {\bibinfo {author} {\bibfnamefont {D.}~\bibnamefont
  {Aloni}}, \bibinfo {author} {\bibfnamefont {K.}~\bibnamefont {Blum}},
  \bibinfo {author} {\bibfnamefont {A.}~\bibnamefont {Dery}}, \bibinfo {author}
  {\bibfnamefont {A.}~\bibnamefont {Efrati}}, \ and\ \bibinfo {author}
  {\bibfnamefont {Y.}~\bibnamefont {Nir}},\ }\href@noop {} {\  (\bibinfo {year}
  {2015})},\ \Eprint {http://arxiv.org/abs/1512.05778} {arXiv:1512.05778
  [hep-ph]} \BibitemShut {NoStop}%
\bibitem [{\citenamefont {Falkowski}\ \emph {et~al.}(2015)\citenamefont
  {Falkowski}, \citenamefont {Slone},\ and\ \citenamefont
  {Volansky}}]{Falkowski:2015swt}%
  \BibitemOpen
  \bibfield  {author} {\bibinfo {author} {\bibfnamefont {A.}~\bibnamefont
  {Falkowski}}, \bibinfo {author} {\bibfnamefont {O.}~\bibnamefont {Slone}}, \
  and\ \bibinfo {author} {\bibfnamefont {T.}~\bibnamefont {Volansky}},\
  }\href@noop {} {\  (\bibinfo {year} {2015})},\ \Eprint
  {http://arxiv.org/abs/1512.05777} {arXiv:1512.05777 [hep-ph]} \BibitemShut
  {NoStop}%
\bibitem [{\citenamefont {Fichet}\ \emph
  {et~al.}(2015{\natexlab{a}})\citenamefont {Fichet}, \citenamefont {von
  Gersdorff},\ and\ \citenamefont {Royon}}]{Fichet:2015vvy}%
  \BibitemOpen
  \bibfield  {author} {\bibinfo {author} {\bibfnamefont {S.}~\bibnamefont
  {Fichet}}, \bibinfo {author} {\bibfnamefont {G.}~\bibnamefont {von
  Gersdorff}}, \ and\ \bibinfo {author} {\bibfnamefont {C.}~\bibnamefont
  {Royon}},\ }\href@noop {} {\  (\bibinfo {year} {2015}{\natexlab{a}})},\
  \Eprint {http://arxiv.org/abs/1512.05751} {arXiv:1512.05751 [hep-ph]}
  \BibitemShut {NoStop}%
\bibitem [{\citenamefont {Csaki}\ \emph {et~al.}(2015)\citenamefont {Csaki},
  \citenamefont {Hubisz},\ and\ \citenamefont {Terning}}]{Csaki:2015vek}%
  \BibitemOpen
  \bibfield  {author} {\bibinfo {author} {\bibfnamefont {C.}~\bibnamefont
  {Csaki}}, \bibinfo {author} {\bibfnamefont {J.}~\bibnamefont {Hubisz}}, \
  and\ \bibinfo {author} {\bibfnamefont {J.}~\bibnamefont {Terning}},\
  }\href@noop {} {\  (\bibinfo {year} {2015})},\ \Eprint
  {http://arxiv.org/abs/1512.05776} {arXiv:1512.05776 [hep-ph]} \BibitemShut
  {NoStop}%
\bibitem [{\citenamefont {Chakrabortty}\ \emph {et~al.}(2015)\citenamefont
  {Chakrabortty}, \citenamefont {Choudhury}, \citenamefont {Ghosh},
  \citenamefont {Mondal},\ and\ \citenamefont
  {Srivastava}}]{Chakrabortty:2015hff}%
  \BibitemOpen
  \bibfield  {author} {\bibinfo {author} {\bibfnamefont {J.}~\bibnamefont
  {Chakrabortty}}, \bibinfo {author} {\bibfnamefont {A.}~\bibnamefont
  {Choudhury}}, \bibinfo {author} {\bibfnamefont {P.}~\bibnamefont {Ghosh}},
  \bibinfo {author} {\bibfnamefont {S.}~\bibnamefont {Mondal}}, \ and\ \bibinfo
  {author} {\bibfnamefont {T.}~\bibnamefont {Srivastava}},\ }\href@noop {} {\
  (\bibinfo {year} {2015})},\ \Eprint {http://arxiv.org/abs/1512.05767}
  {arXiv:1512.05767 [hep-ph]} \BibitemShut {NoStop}%
\bibitem [{\citenamefont {Bian}\ \emph {et~al.}(2015)\citenamefont {Bian},
  \citenamefont {Chen}, \citenamefont {Liu},\ and\ \citenamefont
  {Shu}}]{Bian:2015kjt}%
  \BibitemOpen
  \bibfield  {author} {\bibinfo {author} {\bibfnamefont {L.}~\bibnamefont
  {Bian}}, \bibinfo {author} {\bibfnamefont {N.}~\bibnamefont {Chen}}, \bibinfo
  {author} {\bibfnamefont {D.}~\bibnamefont {Liu}}, \ and\ \bibinfo {author}
  {\bibfnamefont {J.}~\bibnamefont {Shu}},\ }\href@noop {} {\  (\bibinfo {year}
  {2015})},\ \Eprint {http://arxiv.org/abs/1512.05759} {arXiv:1512.05759
  [hep-ph]} \BibitemShut {NoStop}%
\bibitem [{\citenamefont {Curtin}\ and\ \citenamefont
  {Verhaaren}(2015)}]{Curtin:2015jcv}%
  \BibitemOpen
  \bibfield  {author} {\bibinfo {author} {\bibfnamefont {D.}~\bibnamefont
  {Curtin}}\ and\ \bibinfo {author} {\bibfnamefont {C.~B.}\ \bibnamefont
  {Verhaaren}},\ }\href@noop {} {\  (\bibinfo {year} {2015})},\ \Eprint
  {http://arxiv.org/abs/1512.05753} {arXiv:1512.05753 [hep-ph]} \BibitemShut
  {NoStop}%
\bibitem [{\citenamefont {Chao}\ \emph {et~al.}(2015)\citenamefont {Chao},
  \citenamefont {Huo},\ and\ \citenamefont {Yu}}]{Chao:2015ttq}%
  \BibitemOpen
  \bibfield  {author} {\bibinfo {author} {\bibfnamefont {W.}~\bibnamefont
  {Chao}}, \bibinfo {author} {\bibfnamefont {R.}~\bibnamefont {Huo}}, \ and\
  \bibinfo {author} {\bibfnamefont {J.-H.}\ \bibnamefont {Yu}},\ }\href@noop {}
  {\  (\bibinfo {year} {2015})},\ \Eprint {http://arxiv.org/abs/1512.05738}
  {arXiv:1512.05738 [hep-ph]} \BibitemShut {NoStop}%
\bibitem [{\citenamefont {Demidov}\ and\ \citenamefont
  {Gorbunov}(2015)}]{Demidov:2015zqn}%
  \BibitemOpen
  \bibfield  {author} {\bibinfo {author} {\bibfnamefont {S.~V.}\ \bibnamefont
  {Demidov}}\ and\ \bibinfo {author} {\bibfnamefont {D.~S.}\ \bibnamefont
  {Gorbunov}},\ }\href@noop {} {\  (\bibinfo {year} {2015})},\ \Eprint
  {http://arxiv.org/abs/1512.05723} {arXiv:1512.05723 [hep-ph]} \BibitemShut
  {NoStop}%
\bibitem [{\citenamefont {No}\ \emph {et~al.}(2015)\citenamefont {No},
  \citenamefont {Sanz},\ and\ \citenamefont {Setford}}]{No:2015bsn}%
  \BibitemOpen
  \bibfield  {author} {\bibinfo {author} {\bibfnamefont {J.~M.}\ \bibnamefont
  {No}}, \bibinfo {author} {\bibfnamefont {V.}~\bibnamefont {Sanz}}, \ and\
  \bibinfo {author} {\bibfnamefont {J.}~\bibnamefont {Setford}},\ }\href@noop
  {} {\  (\bibinfo {year} {2015})},\ \Eprint {http://arxiv.org/abs/1512.05700}
  {arXiv:1512.05700 [hep-ph]} \BibitemShut {NoStop}%
\bibitem [{\citenamefont {Becirevic}\ \emph {et~al.}(2015)\citenamefont
  {Becirevic}, \citenamefont {Bertuzzo}, \citenamefont {Sumensari},\ and\
  \citenamefont {Funchal}}]{Becirevic:2015fmu}%
  \BibitemOpen
  \bibfield  {author} {\bibinfo {author} {\bibfnamefont {D.}~\bibnamefont
  {Becirevic}}, \bibinfo {author} {\bibfnamefont {E.}~\bibnamefont {Bertuzzo}},
  \bibinfo {author} {\bibfnamefont {O.}~\bibnamefont {Sumensari}}, \ and\
  \bibinfo {author} {\bibfnamefont {R.~Z.}\ \bibnamefont {Funchal}},\
  }\href@noop {} {\  (\bibinfo {year} {2015})},\ \Eprint
  {http://arxiv.org/abs/1512.05623} {arXiv:1512.05623 [hep-ph]} \BibitemShut
  {NoStop}%
\bibitem [{\citenamefont {Martinez}\ \emph {et~al.}(2015)\citenamefont
  {Martinez}, \citenamefont {Ochoa},\ and\ \citenamefont
  {Sierra}}]{Martinez:2015kmn}%
  \BibitemOpen
  \bibfield  {author} {\bibinfo {author} {\bibfnamefont {R.}~\bibnamefont
  {Martinez}}, \bibinfo {author} {\bibfnamefont {F.}~\bibnamefont {Ochoa}}, \
  and\ \bibinfo {author} {\bibfnamefont {C.~F.}\ \bibnamefont {Sierra}},\
  }\href@noop {} {\  (\bibinfo {year} {2015})},\ \Eprint
  {http://arxiv.org/abs/1512.05617} {arXiv:1512.05617 [hep-ph]} \BibitemShut
  {NoStop}%
\bibitem [{\citenamefont {Agrawal}\ \emph {et~al.}(2015)\citenamefont
  {Agrawal}, \citenamefont {Fan}, \citenamefont {Heidenreich}, \citenamefont
  {Reece},\ and\ \citenamefont {Strassler}}]{Agrawal:2015dbf}%
  \BibitemOpen
  \bibfield  {author} {\bibinfo {author} {\bibfnamefont {P.}~\bibnamefont
  {Agrawal}}, \bibinfo {author} {\bibfnamefont {J.}~\bibnamefont {Fan}},
  \bibinfo {author} {\bibfnamefont {B.}~\bibnamefont {Heidenreich}}, \bibinfo
  {author} {\bibfnamefont {M.}~\bibnamefont {Reece}}, \ and\ \bibinfo {author}
  {\bibfnamefont {M.}~\bibnamefont {Strassler}},\ }\href@noop {} {\  (\bibinfo
  {year} {2015})},\ \Eprint {http://arxiv.org/abs/1512.05775} {arXiv:1512.05775
  [hep-ph]} \BibitemShut {NoStop}%
\bibitem [{\citenamefont {Ahmed}\ \emph {et~al.}(2015)\citenamefont {Ahmed},
  \citenamefont {Dillon}, \citenamefont {Grzadkowski}, \citenamefont {Gunion},\
  and\ \citenamefont {Jiang}}]{Ahmed:2015uqt}%
  \BibitemOpen
  \bibfield  {author} {\bibinfo {author} {\bibfnamefont {A.}~\bibnamefont
  {Ahmed}}, \bibinfo {author} {\bibfnamefont {B.~M.}\ \bibnamefont {Dillon}},
  \bibinfo {author} {\bibfnamefont {B.}~\bibnamefont {Grzadkowski}}, \bibinfo
  {author} {\bibfnamefont {J.~F.}\ \bibnamefont {Gunion}}, \ and\ \bibinfo
  {author} {\bibfnamefont {Y.}~\bibnamefont {Jiang}},\ }\href@noop {} {\
  (\bibinfo {year} {2015})},\ \Eprint {http://arxiv.org/abs/1512.05771}
  {arXiv:1512.05771 [hep-ph]} \BibitemShut {NoStop}%
\bibitem [{\citenamefont {Cox}\ \emph {et~al.}(2015)\citenamefont {Cox},
  \citenamefont {Medina}, \citenamefont {Ray},\ and\ \citenamefont
  {Spray}}]{Cox:2015ckc}%
  \BibitemOpen
  \bibfield  {author} {\bibinfo {author} {\bibfnamefont {P.}~\bibnamefont
  {Cox}}, \bibinfo {author} {\bibfnamefont {A.~D.}\ \bibnamefont {Medina}},
  \bibinfo {author} {\bibfnamefont {T.~S.}\ \bibnamefont {Ray}}, \ and\
  \bibinfo {author} {\bibfnamefont {A.}~\bibnamefont {Spray}},\ }\href@noop {}
  {\  (\bibinfo {year} {2015})},\ \Eprint {http://arxiv.org/abs/1512.05618}
  {arXiv:1512.05618 [hep-ph]} \BibitemShut {NoStop}%
\bibitem [{\citenamefont {Kobakhidze}\ \emph {et~al.}(2015)\citenamefont
  {Kobakhidze}, \citenamefont {Wang}, \citenamefont {Wu}, \citenamefont
  {Yang},\ and\ \citenamefont {Zhang}}]{Kobakhidze:2015ldh}%
  \BibitemOpen
  \bibfield  {author} {\bibinfo {author} {\bibfnamefont {A.}~\bibnamefont
  {Kobakhidze}}, \bibinfo {author} {\bibfnamefont {F.}~\bibnamefont {Wang}},
  \bibinfo {author} {\bibfnamefont {L.}~\bibnamefont {Wu}}, \bibinfo {author}
  {\bibfnamefont {J.~M.}\ \bibnamefont {Yang}}, \ and\ \bibinfo {author}
  {\bibfnamefont {M.}~\bibnamefont {Zhang}},\ }\href@noop {} {\  (\bibinfo
  {year} {2015})},\ \Eprint {http://arxiv.org/abs/1512.05585} {arXiv:1512.05585
  [hep-ph]} \BibitemShut {NoStop}%
\bibitem [{\citenamefont {Matsuzaki}\ and\ \citenamefont
  {Yamawaki}(2015)}]{Matsuzaki:2015che}%
  \BibitemOpen
  \bibfield  {author} {\bibinfo {author} {\bibfnamefont {S.}~\bibnamefont
  {Matsuzaki}}\ and\ \bibinfo {author} {\bibfnamefont {K.}~\bibnamefont
  {Yamawaki}},\ }\href@noop {} {\  (\bibinfo {year} {2015})},\ \Eprint
  {http://arxiv.org/abs/1512.05564} {arXiv:1512.05564 [hep-ph]} \BibitemShut
  {NoStop}%
\bibitem [{\citenamefont {Cao}\ \emph {et~al.}(2015{\natexlab{a}})\citenamefont
  {Cao}, \citenamefont {Liu}, \citenamefont {Xie}, \citenamefont {Yan},\ and\
  \citenamefont {Zhang}}]{Cao:2015pto}%
  \BibitemOpen
  \bibfield  {author} {\bibinfo {author} {\bibfnamefont {Q.-H.}\ \bibnamefont
  {Cao}}, \bibinfo {author} {\bibfnamefont {Y.}~\bibnamefont {Liu}}, \bibinfo
  {author} {\bibfnamefont {K.-P.}\ \bibnamefont {Xie}}, \bibinfo {author}
  {\bibfnamefont {B.}~\bibnamefont {Yan}}, \ and\ \bibinfo {author}
  {\bibfnamefont {D.-M.}\ \bibnamefont {Zhang}},\ }\href@noop {} {\  (\bibinfo
  {year} {2015}{\natexlab{a}})},\ \Eprint {http://arxiv.org/abs/1512.05542}
  {arXiv:1512.05542 [hep-ph]} \BibitemShut {NoStop}%
\bibitem [{\citenamefont {Benbrik}\ \emph {et~al.}(2015)\citenamefont
  {Benbrik}, \citenamefont {Chen},\ and\ \citenamefont
  {Nomura}}]{Benbrik:2015fyz}%
  \BibitemOpen
  \bibfield  {author} {\bibinfo {author} {\bibfnamefont {R.}~\bibnamefont
  {Benbrik}}, \bibinfo {author} {\bibfnamefont {C.-H.}\ \bibnamefont {Chen}}, \
  and\ \bibinfo {author} {\bibfnamefont {T.}~\bibnamefont {Nomura}},\
  }\href@noop {} {\  (\bibinfo {year} {2015})},\ \Eprint
  {http://arxiv.org/abs/1512.06028} {arXiv:1512.06028 [hep-ph]} \BibitemShut
  {NoStop}%
\bibitem [{\citenamefont {Kim}\ \emph {et~al.}(2015{\natexlab{a}})\citenamefont
  {Kim}, \citenamefont {Reuter}, \citenamefont {Rolbiecki},\ and\ \citenamefont
  {de~Austri}}]{Kim:2015ron}%
  \BibitemOpen
  \bibfield  {author} {\bibinfo {author} {\bibfnamefont {J.~S.}\ \bibnamefont
  {Kim}}, \bibinfo {author} {\bibfnamefont {J.}~\bibnamefont {Reuter}},
  \bibinfo {author} {\bibfnamefont {K.}~\bibnamefont {Rolbiecki}}, \ and\
  \bibinfo {author} {\bibfnamefont {R.~R.}\ \bibnamefont {de~Austri}},\
  }\href@noop {} {\  (\bibinfo {year} {2015}{\natexlab{a}})},\ \Eprint
  {http://arxiv.org/abs/1512.06083} {arXiv:1512.06083 [hep-ph]} \BibitemShut
  {NoStop}%
\bibitem [{\citenamefont {Gabrielli}\ \emph {et~al.}(2015)\citenamefont
  {Gabrielli}, \citenamefont {Kannike}, \citenamefont {Mele}, \citenamefont
  {Raidal}, \citenamefont {Spethmann},\ and\ \citenamefont
  {VeermŠe}}]{Gabrielli:2015dhk}%
  \BibitemOpen
  \bibfield  {author} {\bibinfo {author} {\bibfnamefont {E.}~\bibnamefont
  {Gabrielli}}, \bibinfo {author} {\bibfnamefont {K.}~\bibnamefont {Kannike}},
  \bibinfo {author} {\bibfnamefont {B.}~\bibnamefont {Mele}}, \bibinfo {author}
  {\bibfnamefont {M.}~\bibnamefont {Raidal}}, \bibinfo {author} {\bibfnamefont
  {C.}~\bibnamefont {Spethmann}}, \ and\ \bibinfo {author} {\bibfnamefont
  {H.}~\bibnamefont {VeermŠe}},\ }\href@noop {} {\  (\bibinfo {year} {2015})},\
  \Eprint {http://arxiv.org/abs/1512.05961} {arXiv:1512.05961 [hep-ph]}
  \BibitemShut {NoStop}%
\bibitem [{\citenamefont {Alves}\ \emph {et~al.}(2015)\citenamefont {Alves},
  \citenamefont {Dias},\ and\ \citenamefont {Sinha}}]{Alves:2015jgx}%
  \BibitemOpen
  \bibfield  {author} {\bibinfo {author} {\bibfnamefont {A.}~\bibnamefont
  {Alves}}, \bibinfo {author} {\bibfnamefont {A.~G.}\ \bibnamefont {Dias}}, \
  and\ \bibinfo {author} {\bibfnamefont {K.}~\bibnamefont {Sinha}},\
  }\href@noop {} {\  (\bibinfo {year} {2015})},\ \Eprint
  {http://arxiv.org/abs/1512.06091} {arXiv:1512.06091 [hep-ph]} \BibitemShut
  {NoStop}%
\bibitem [{\citenamefont {Bernon}\ and\ \citenamefont
  {Smith}(2015)}]{Bernon:2015abk}%
  \BibitemOpen
  \bibfield  {author} {\bibinfo {author} {\bibfnamefont {J.}~\bibnamefont
  {Bernon}}\ and\ \bibinfo {author} {\bibfnamefont {C.}~\bibnamefont {Smith}},\
  }\href@noop {} {\  (\bibinfo {year} {2015})},\ \Eprint
  {http://arxiv.org/abs/1512.06113} {arXiv:1512.06113 [hep-ph]} \BibitemShut
  {NoStop}%
\bibitem [{\citenamefont {Dhuria}\ and\ \citenamefont
  {Goswami}(2015)}]{Dhuria:2015ufo}%
  \BibitemOpen
  \bibfield  {author} {\bibinfo {author} {\bibfnamefont {M.}~\bibnamefont
  {Dhuria}}\ and\ \bibinfo {author} {\bibfnamefont {G.}~\bibnamefont
  {Goswami}},\ }\href@noop {} {\  (\bibinfo {year} {2015})},\ \Eprint
  {http://arxiv.org/abs/1512.06782} {arXiv:1512.06782 [hep-ph]} \BibitemShut
  {NoStop}%
\bibitem [{\citenamefont {Han}\ \emph {et~al.}(2015{\natexlab{a}})\citenamefont
  {Han}, \citenamefont {Lee}, \citenamefont {Park},\ and\ \citenamefont
  {Sanz}}]{Han:2015cty}%
  \BibitemOpen
  \bibfield  {author} {\bibinfo {author} {\bibfnamefont {C.}~\bibnamefont
  {Han}}, \bibinfo {author} {\bibfnamefont {H.~M.}\ \bibnamefont {Lee}},
  \bibinfo {author} {\bibfnamefont {M.}~\bibnamefont {Park}}, \ and\ \bibinfo
  {author} {\bibfnamefont {V.}~\bibnamefont {Sanz}},\ }\href@noop {} {\
  (\bibinfo {year} {2015}{\natexlab{a}})},\ \Eprint
  {http://arxiv.org/abs/1512.06376} {arXiv:1512.06376 [hep-ph]} \BibitemShut
  {NoStop}%
\bibitem [{\citenamefont {Han}\ \emph {et~al.}(2015{\natexlab{b}})\citenamefont
  {Han}, \citenamefont {Wang},\ and\ \citenamefont {Zheng}}]{Han:2015dlp}%
  \BibitemOpen
  \bibfield  {author} {\bibinfo {author} {\bibfnamefont {H.}~\bibnamefont
  {Han}}, \bibinfo {author} {\bibfnamefont {S.}~\bibnamefont {Wang}}, \ and\
  \bibinfo {author} {\bibfnamefont {S.}~\bibnamefont {Zheng}},\ }\href@noop {}
  {\  (\bibinfo {year} {2015}{\natexlab{b}})},\ \Eprint
  {http://arxiv.org/abs/1512.06562} {arXiv:1512.06562 [hep-ph]} \BibitemShut
  {NoStop}%
\bibitem [{\citenamefont {Luo}\ \emph {et~al.}(2015)\citenamefont {Luo},
  \citenamefont {Wang}, \citenamefont {Xu}, \citenamefont {Zhang},\ and\
  \citenamefont {Zhu}}]{Luo:2015yio}%
  \BibitemOpen
  \bibfield  {author} {\bibinfo {author} {\bibfnamefont {M.-x.}\ \bibnamefont
  {Luo}}, \bibinfo {author} {\bibfnamefont {K.}~\bibnamefont {Wang}}, \bibinfo
  {author} {\bibfnamefont {T.}~\bibnamefont {Xu}}, \bibinfo {author}
  {\bibfnamefont {L.}~\bibnamefont {Zhang}}, \ and\ \bibinfo {author}
  {\bibfnamefont {G.}~\bibnamefont {Zhu}},\ }\href@noop {} {\  (\bibinfo {year}
  {2015})},\ \Eprint {http://arxiv.org/abs/1512.06670} {arXiv:1512.06670
  [hep-ph]} \BibitemShut {NoStop}%
\bibitem [{\citenamefont {Chang}\ \emph {et~al.}(2015)\citenamefont {Chang},
  \citenamefont {Cheung},\ and\ \citenamefont {Lu}}]{Chang:2015sdy}%
  \BibitemOpen
  \bibfield  {author} {\bibinfo {author} {\bibfnamefont {J.}~\bibnamefont
  {Chang}}, \bibinfo {author} {\bibfnamefont {K.}~\bibnamefont {Cheung}}, \
  and\ \bibinfo {author} {\bibfnamefont {C.-T.}\ \bibnamefont {Lu}},\
  }\href@noop {} {\  (\bibinfo {year} {2015})},\ \Eprint
  {http://arxiv.org/abs/1512.06671} {arXiv:1512.06671 [hep-ph]} \BibitemShut
  {NoStop}%
\bibitem [{\citenamefont {Bardhan}\ \emph {et~al.}(2015)\citenamefont
  {Bardhan}, \citenamefont {Bhatia}, \citenamefont {Chakraborty}, \citenamefont
  {Maitra}, \citenamefont {Raychaudhuri},\ and\ \citenamefont
  {Samui}}]{Bardhan:2015hcr}%
  \BibitemOpen
  \bibfield  {author} {\bibinfo {author} {\bibfnamefont {D.}~\bibnamefont
  {Bardhan}}, \bibinfo {author} {\bibfnamefont {D.}~\bibnamefont {Bhatia}},
  \bibinfo {author} {\bibfnamefont {A.}~\bibnamefont {Chakraborty}}, \bibinfo
  {author} {\bibfnamefont {U.}~\bibnamefont {Maitra}}, \bibinfo {author}
  {\bibfnamefont {S.}~\bibnamefont {Raychaudhuri}}, \ and\ \bibinfo {author}
  {\bibfnamefont {T.}~\bibnamefont {Samui}},\ }\href@noop {} {\  (\bibinfo
  {year} {2015})},\ \Eprint {http://arxiv.org/abs/1512.06674} {arXiv:1512.06674
  [hep-ph]} \BibitemShut {NoStop}%
\bibitem [{\citenamefont {Feng}\ \emph {et~al.}(2015)\citenamefont {Feng},
  \citenamefont {Li}, \citenamefont {Zhang},\ and\ \citenamefont
  {Zhao}}]{Feng:2015wil}%
  \BibitemOpen
  \bibfield  {author} {\bibinfo {author} {\bibfnamefont {T.-F.}\ \bibnamefont
  {Feng}}, \bibinfo {author} {\bibfnamefont {X.-Q.}\ \bibnamefont {Li}},
  \bibinfo {author} {\bibfnamefont {H.-B.}\ \bibnamefont {Zhang}}, \ and\
  \bibinfo {author} {\bibfnamefont {S.-M.}\ \bibnamefont {Zhao}},\ }\href@noop
  {} {\  (\bibinfo {year} {2015})},\ \Eprint {http://arxiv.org/abs/1512.06696}
  {arXiv:1512.06696 [hep-ph]} \BibitemShut {NoStop}%
\bibitem [{\citenamefont {Barducci}\ \emph {et~al.}(2015)\citenamefont
  {Barducci}, \citenamefont {Goudelis}, \citenamefont {Kulkarni},\ and\
  \citenamefont {Sengupta}}]{Barducci:2015gtd}%
  \BibitemOpen
  \bibfield  {author} {\bibinfo {author} {\bibfnamefont {D.}~\bibnamefont
  {Barducci}}, \bibinfo {author} {\bibfnamefont {A.}~\bibnamefont {Goudelis}},
  \bibinfo {author} {\bibfnamefont {S.}~\bibnamefont {Kulkarni}}, \ and\
  \bibinfo {author} {\bibfnamefont {D.}~\bibnamefont {Sengupta}},\ }\href@noop
  {} {\  (\bibinfo {year} {2015})},\ \Eprint {http://arxiv.org/abs/1512.06842}
  {arXiv:1512.06842 [hep-ph]} \BibitemShut {NoStop}%
\bibitem [{\citenamefont {Chao}(2015{\natexlab{a}})}]{Chao:2015nsm}%
  \BibitemOpen
  \bibfield  {author} {\bibinfo {author} {\bibfnamefont {W.}~\bibnamefont
  {Chao}},\ }\href@noop {} {\  (\bibinfo {year} {2015}{\natexlab{a}})},\
  \Eprint {http://arxiv.org/abs/1512.06297} {arXiv:1512.06297 [hep-ph]}
  \BibitemShut {NoStop}%
\bibitem [{\citenamefont {Chakraborty}\ and\ \citenamefont
  {Kundu}(2015)}]{Chakraborty:2015jvs}%
  \BibitemOpen
  \bibfield  {author} {\bibinfo {author} {\bibfnamefont {I.}~\bibnamefont
  {Chakraborty}}\ and\ \bibinfo {author} {\bibfnamefont {A.}~\bibnamefont
  {Kundu}},\ }\href@noop {} {\  (\bibinfo {year} {2015})},\ \Eprint
  {http://arxiv.org/abs/1512.06508} {arXiv:1512.06508 [hep-ph]} \BibitemShut
  {NoStop}%
\bibitem [{\citenamefont {Ding}\ \emph {et~al.}(2015)\citenamefont {Ding},
  \citenamefont {Huang}, \citenamefont {Li},\ and\ \citenamefont
  {Zhu}}]{Ding:2015rxx}%
  \BibitemOpen
  \bibfield  {author} {\bibinfo {author} {\bibfnamefont {R.}~\bibnamefont
  {Ding}}, \bibinfo {author} {\bibfnamefont {L.}~\bibnamefont {Huang}},
  \bibinfo {author} {\bibfnamefont {T.}~\bibnamefont {Li}}, \ and\ \bibinfo
  {author} {\bibfnamefont {B.}~\bibnamefont {Zhu}},\ }\href@noop {} {\
  (\bibinfo {year} {2015})},\ \Eprint {http://arxiv.org/abs/1512.06560}
  {arXiv:1512.06560 [hep-ph]} \BibitemShut {NoStop}%
\bibitem [{\citenamefont {Han}\ and\ \citenamefont {Wang}(2015)}]{Han:2015qqj}%
  \BibitemOpen
  \bibfield  {author} {\bibinfo {author} {\bibfnamefont {X.-F.}\ \bibnamefont
  {Han}}\ and\ \bibinfo {author} {\bibfnamefont {L.}~\bibnamefont {Wang}},\
  }\href@noop {} {\  (\bibinfo {year} {2015})},\ \Eprint
  {http://arxiv.org/abs/1512.06587} {arXiv:1512.06587 [hep-ph]} \BibitemShut
  {NoStop}%
\bibitem [{\citenamefont {Wang}\ \emph
  {et~al.}(2015{\natexlab{a}})\citenamefont {Wang}, \citenamefont {Wu},
  \citenamefont {Yang},\ and\ \citenamefont {Zhang}}]{Wang:2015kuj}%
  \BibitemOpen
  \bibfield  {author} {\bibinfo {author} {\bibfnamefont {F.}~\bibnamefont
  {Wang}}, \bibinfo {author} {\bibfnamefont {L.}~\bibnamefont {Wu}}, \bibinfo
  {author} {\bibfnamefont {J.~M.}\ \bibnamefont {Yang}}, \ and\ \bibinfo
  {author} {\bibfnamefont {M.}~\bibnamefont {Zhang}},\ }\href@noop {} {\
  (\bibinfo {year} {2015}{\natexlab{a}})},\ \Eprint
  {http://arxiv.org/abs/1512.06715} {arXiv:1512.06715 [hep-ph]} \BibitemShut
  {NoStop}%
\bibitem [{\citenamefont {Cao}\ \emph {et~al.}(2015{\natexlab{b}})\citenamefont
  {Cao}, \citenamefont {Han}, \citenamefont {Shang}, \citenamefont {Su},
  \citenamefont {Yang},\ and\ \citenamefont {Zhang}}]{Cao:2015twy}%
  \BibitemOpen
  \bibfield  {author} {\bibinfo {author} {\bibfnamefont {J.}~\bibnamefont
  {Cao}}, \bibinfo {author} {\bibfnamefont {C.}~\bibnamefont {Han}}, \bibinfo
  {author} {\bibfnamefont {L.}~\bibnamefont {Shang}}, \bibinfo {author}
  {\bibfnamefont {W.}~\bibnamefont {Su}}, \bibinfo {author} {\bibfnamefont
  {J.~M.}\ \bibnamefont {Yang}}, \ and\ \bibinfo {author} {\bibfnamefont
  {Y.}~\bibnamefont {Zhang}},\ }\href@noop {} {\  (\bibinfo {year}
  {2015}{\natexlab{b}})},\ \Eprint {http://arxiv.org/abs/1512.06728}
  {arXiv:1512.06728 [hep-ph]} \BibitemShut {NoStop}%
\bibitem [{\citenamefont {Huang}\ \emph
  {et~al.}(2015{\natexlab{a}})\citenamefont {Huang}, \citenamefont {Li},
  \citenamefont {Liu},\ and\ \citenamefont {Wang}}]{Huang:2015evq}%
  \BibitemOpen
  \bibfield  {author} {\bibinfo {author} {\bibfnamefont {F.~P.}\ \bibnamefont
  {Huang}}, \bibinfo {author} {\bibfnamefont {C.~S.}\ \bibnamefont {Li}},
  \bibinfo {author} {\bibfnamefont {Z.~L.}\ \bibnamefont {Liu}}, \ and\
  \bibinfo {author} {\bibfnamefont {Y.}~\bibnamefont {Wang}},\ }\href@noop {}
  {\  (\bibinfo {year} {2015}{\natexlab{a}})},\ \Eprint
  {http://arxiv.org/abs/1512.06732} {arXiv:1512.06732 [hep-ph]} \BibitemShut
  {NoStop}%
\bibitem [{\citenamefont {Heckman}(2015)}]{Heckman:2015kqk}%
  \BibitemOpen
  \bibfield  {author} {\bibinfo {author} {\bibfnamefont {J.~J.}\ \bibnamefont
  {Heckman}},\ }\href@noop {} {\  (\bibinfo {year} {2015})},\ \Eprint
  {http://arxiv.org/abs/1512.06773} {arXiv:1512.06773 [hep-ph]} \BibitemShut
  {NoStop}%
\bibitem [{\citenamefont {Bi}\ \emph {et~al.}(2015{\natexlab{a}})\citenamefont
  {Bi}, \citenamefont {Xiang}, \citenamefont {Yin},\ and\ \citenamefont
  {Yu}}]{Bi:2015uqd}%
  \BibitemOpen
  \bibfield  {author} {\bibinfo {author} {\bibfnamefont {X.-J.}\ \bibnamefont
  {Bi}}, \bibinfo {author} {\bibfnamefont {Q.-F.}\ \bibnamefont {Xiang}},
  \bibinfo {author} {\bibfnamefont {P.-F.}\ \bibnamefont {Yin}}, \ and\
  \bibinfo {author} {\bibfnamefont {Z.-H.}\ \bibnamefont {Yu}},\ }\href@noop {}
  {\  (\bibinfo {year} {2015}{\natexlab{a}})},\ \Eprint
  {http://arxiv.org/abs/1512.06787} {arXiv:1512.06787 [hep-ph]} \BibitemShut
  {NoStop}%
\bibitem [{\citenamefont {Kim}\ \emph {et~al.}(2015{\natexlab{b}})\citenamefont
  {Kim}, \citenamefont {Rolbiecki},\ and\ \citenamefont
  {de~Austri}}]{Kim:2015ksf}%
  \BibitemOpen
  \bibfield  {author} {\bibinfo {author} {\bibfnamefont {J.~S.}\ \bibnamefont
  {Kim}}, \bibinfo {author} {\bibfnamefont {K.}~\bibnamefont {Rolbiecki}}, \
  and\ \bibinfo {author} {\bibfnamefont {R.~R.}\ \bibnamefont {de~Austri}},\
  }\href@noop {} {\  (\bibinfo {year} {2015}{\natexlab{b}})},\ \Eprint
  {http://arxiv.org/abs/1512.06797} {arXiv:1512.06797 [hep-ph]} \BibitemShut
  {NoStop}%
\bibitem [{\citenamefont {Cline}\ and\ \citenamefont
  {Liu}(2015)}]{Cline:2015msi}%
  \BibitemOpen
  \bibfield  {author} {\bibinfo {author} {\bibfnamefont {J.~M.}\ \bibnamefont
  {Cline}}\ and\ \bibinfo {author} {\bibfnamefont {Z.}~\bibnamefont {Liu}},\
  }\href@noop {} {\  (\bibinfo {year} {2015})},\ \Eprint
  {http://arxiv.org/abs/1512.06827} {arXiv:1512.06827 [hep-ph]} \BibitemShut
  {NoStop}%
\bibitem [{\citenamefont {Bauer}\ and\ \citenamefont
  {Neubert}(2015)}]{Bauer:2015boy}%
  \BibitemOpen
  \bibfield  {author} {\bibinfo {author} {\bibfnamefont {M.}~\bibnamefont
  {Bauer}}\ and\ \bibinfo {author} {\bibfnamefont {M.}~\bibnamefont
  {Neubert}},\ }\href@noop {} {\  (\bibinfo {year} {2015})},\ \Eprint
  {http://arxiv.org/abs/1512.06828} {arXiv:1512.06828 [hep-ph]} \BibitemShut
  {NoStop}%
\bibitem [{\citenamefont {Chala}\ \emph {et~al.}(2015)\citenamefont {Chala},
  \citenamefont {Duerr}, \citenamefont {Kahlhoefer},\ and\ \citenamefont
  {Schmidt-Hoberg}}]{Chala:2015cev}%
  \BibitemOpen
  \bibfield  {author} {\bibinfo {author} {\bibfnamefont {M.}~\bibnamefont
  {Chala}}, \bibinfo {author} {\bibfnamefont {M.}~\bibnamefont {Duerr}},
  \bibinfo {author} {\bibfnamefont {F.}~\bibnamefont {Kahlhoefer}}, \ and\
  \bibinfo {author} {\bibfnamefont {K.}~\bibnamefont {Schmidt-Hoberg}},\
  }\href@noop {} {\  (\bibinfo {year} {2015})},\ \Eprint
  {http://arxiv.org/abs/1512.06833} {arXiv:1512.06833 [hep-ph]} \BibitemShut
  {NoStop}%
\bibitem [{\citenamefont {de~Blas}\ \emph {et~al.}(2015)\citenamefont
  {de~Blas}, \citenamefont {Santiago},\ and\ \citenamefont
  {Vega-Morales}}]{deBlas:2015hlv}%
  \BibitemOpen
  \bibfield  {author} {\bibinfo {author} {\bibfnamefont {J.}~\bibnamefont
  {de~Blas}}, \bibinfo {author} {\bibfnamefont {J.}~\bibnamefont {Santiago}}, \
  and\ \bibinfo {author} {\bibfnamefont {R.}~\bibnamefont {Vega-Morales}},\
  }\href@noop {} {\  (\bibinfo {year} {2015})},\ \Eprint
  {http://arxiv.org/abs/1512.07229} {arXiv:1512.07229 [hep-ph]} \BibitemShut
  {NoStop}%
\bibitem [{\citenamefont {Boucenna}\ \emph {et~al.}(2015)\citenamefont
  {Boucenna}, \citenamefont {Morisi},\ and\ \citenamefont
  {Vicente}}]{Boucenna:2015pav}%
  \BibitemOpen
  \bibfield  {author} {\bibinfo {author} {\bibfnamefont {S.~M.}\ \bibnamefont
  {Boucenna}}, \bibinfo {author} {\bibfnamefont {S.}~\bibnamefont {Morisi}}, \
  and\ \bibinfo {author} {\bibfnamefont {A.}~\bibnamefont {Vicente}},\
  }\href@noop {} {\  (\bibinfo {year} {2015})},\ \Eprint
  {http://arxiv.org/abs/1512.06878} {arXiv:1512.06878 [hep-ph]} \BibitemShut
  {NoStop}%
\bibitem [{\citenamefont {Murphy}(2015)}]{Murphy:2015kag}%
  \BibitemOpen
  \bibfield  {author} {\bibinfo {author} {\bibfnamefont {C.~W.}\ \bibnamefont
  {Murphy}},\ }\href@noop {} {\  (\bibinfo {year} {2015})},\ \Eprint
  {http://arxiv.org/abs/1512.06976} {arXiv:1512.06976 [hep-ph]} \BibitemShut
  {NoStop}%
\bibitem [{\citenamefont {Hern‡ndez}\ and\ \citenamefont
  {Nisandzic}(2015)}]{Hernandez:2015ywg}%
  \BibitemOpen
  \bibfield  {author} {\bibinfo {author} {\bibfnamefont {A.~E.~C.}\
  \bibnamefont {Hern‡ndez}}\ and\ \bibinfo {author} {\bibfnamefont
  {I.}~\bibnamefont {Nisandzic}},\ }\href@noop {} {\  (\bibinfo {year}
  {2015})},\ \Eprint {http://arxiv.org/abs/1512.07165} {arXiv:1512.07165
  [hep-ph]} \BibitemShut {NoStop}%
\bibitem [{\citenamefont {Dey}\ \emph {et~al.}(2015)\citenamefont {Dey},
  \citenamefont {Mohanty},\ and\ \citenamefont {Tomar}}]{Dey:2015bur}%
  \BibitemOpen
  \bibfield  {author} {\bibinfo {author} {\bibfnamefont {U.~K.}\ \bibnamefont
  {Dey}}, \bibinfo {author} {\bibfnamefont {S.}~\bibnamefont {Mohanty}}, \ and\
  \bibinfo {author} {\bibfnamefont {G.}~\bibnamefont {Tomar}},\ }\href@noop {}
  {\  (\bibinfo {year} {2015})},\ \Eprint {http://arxiv.org/abs/1512.07212}
  {arXiv:1512.07212 [hep-ph]} \BibitemShut {NoStop}%
\bibitem [{\citenamefont {Pelaggi}\ \emph {et~al.}(2015)\citenamefont
  {Pelaggi}, \citenamefont {Strumia},\ and\ \citenamefont
  {Vigiani}}]{Pelaggi:2015knk}%
  \BibitemOpen
  \bibfield  {author} {\bibinfo {author} {\bibfnamefont {G.~M.}\ \bibnamefont
  {Pelaggi}}, \bibinfo {author} {\bibfnamefont {A.}~\bibnamefont {Strumia}}, \
  and\ \bibinfo {author} {\bibfnamefont {E.}~\bibnamefont {Vigiani}},\
  }\href@noop {} {\  (\bibinfo {year} {2015})},\ \Eprint
  {http://arxiv.org/abs/1512.07225} {arXiv:1512.07225 [hep-ph]} \BibitemShut
  {NoStop}%
\bibitem [{\citenamefont {Cao}\ \emph {et~al.}(2015{\natexlab{c}})\citenamefont
  {Cao}, \citenamefont {Chen},\ and\ \citenamefont {Gu}}]{Cao:2015xjz}%
  \BibitemOpen
  \bibfield  {author} {\bibinfo {author} {\bibfnamefont {Q.-H.}\ \bibnamefont
  {Cao}}, \bibinfo {author} {\bibfnamefont {S.-L.}\ \bibnamefont {Chen}}, \
  and\ \bibinfo {author} {\bibfnamefont {P.-H.}\ \bibnamefont {Gu}},\
  }\href@noop {} {\  (\bibinfo {year} {2015}{\natexlab{c}})},\ \Eprint
  {http://arxiv.org/abs/1512.07541} {arXiv:1512.07541 [hep-ph]} \BibitemShut
  {NoStop}%
\bibitem [{\citenamefont {Huang}\ \emph
  {et~al.}(2015{\natexlab{b}})\citenamefont {Huang}, \citenamefont {Tsai},\
  and\ \citenamefont {Yuan}}]{Huang:2015rkj}%
  \BibitemOpen
  \bibfield  {author} {\bibinfo {author} {\bibfnamefont {W.-C.}\ \bibnamefont
  {Huang}}, \bibinfo {author} {\bibfnamefont {Y.-L.~S.}\ \bibnamefont {Tsai}},
  \ and\ \bibinfo {author} {\bibfnamefont {T.-C.}\ \bibnamefont {Yuan}},\
  }\href@noop {} {\  (\bibinfo {year} {2015}{\natexlab{b}})},\ \Eprint
  {http://arxiv.org/abs/1512.07268} {arXiv:1512.07268 [hep-ph]} \BibitemShut
  {NoStop}%
\bibitem [{\citenamefont {Patel}\ and\ \citenamefont
  {Sharma}(2015)}]{Patel:2015ulo}%
  \BibitemOpen
  \bibfield  {author} {\bibinfo {author} {\bibfnamefont {K.~M.}\ \bibnamefont
  {Patel}}\ and\ \bibinfo {author} {\bibfnamefont {P.}~\bibnamefont {Sharma}},\
  }\href@noop {} {\  (\bibinfo {year} {2015})},\ \Eprint
  {http://arxiv.org/abs/1512.07468} {arXiv:1512.07468 [hep-ph]} \BibitemShut
  {NoStop}%
\bibitem [{\citenamefont {Chakraborty}\ \emph {et~al.}(2015)\citenamefont
  {Chakraborty}, \citenamefont {Chakraborty},\ and\ \citenamefont
  {Raychaudhuri}}]{Chakraborty:2015gyj}%
  \BibitemOpen
  \bibfield  {author} {\bibinfo {author} {\bibfnamefont {S.}~\bibnamefont
  {Chakraborty}}, \bibinfo {author} {\bibfnamefont {A.}~\bibnamefont
  {Chakraborty}}, \ and\ \bibinfo {author} {\bibfnamefont {S.}~\bibnamefont
  {Raychaudhuri}},\ }\href@noop {} {\  (\bibinfo {year} {2015})},\ \Eprint
  {http://arxiv.org/abs/1512.07527} {arXiv:1512.07527 [hep-ph]} \BibitemShut
  {NoStop}%
\bibitem [{\citenamefont {Altmannshofer}\ \emph {et~al.}(2015)\citenamefont
  {Altmannshofer}, \citenamefont {Galloway}, \citenamefont {Gori},
  \citenamefont {Kagan}, \citenamefont {Martin},\ and\ \citenamefont
  {Zupan}}]{Altmannshofer:2015xfo}%
  \BibitemOpen
  \bibfield  {author} {\bibinfo {author} {\bibfnamefont {W.}~\bibnamefont
  {Altmannshofer}}, \bibinfo {author} {\bibfnamefont {J.}~\bibnamefont
  {Galloway}}, \bibinfo {author} {\bibfnamefont {S.}~\bibnamefont {Gori}},
  \bibinfo {author} {\bibfnamefont {A.~L.}\ \bibnamefont {Kagan}}, \bibinfo
  {author} {\bibfnamefont {A.}~\bibnamefont {Martin}}, \ and\ \bibinfo {author}
  {\bibfnamefont {J.}~\bibnamefont {Zupan}},\ }\href@noop {} {\  (\bibinfo
  {year} {2015})},\ \Eprint {http://arxiv.org/abs/1512.07616} {arXiv:1512.07616
  [hep-ph]} \BibitemShut {NoStop}%
\bibitem [{\citenamefont {Cveti?}\ \emph {et~al.}(2015)\citenamefont {Cveti?},
  \citenamefont {Halverson},\ and\ \citenamefont {Langacker}}]{Cvetic:2015vit}%
  \BibitemOpen
  \bibfield  {author} {\bibinfo {author} {\bibfnamefont {M.}~\bibnamefont
  {Cveti?}}, \bibinfo {author} {\bibfnamefont {J.}~\bibnamefont {Halverson}}, \
  and\ \bibinfo {author} {\bibfnamefont {P.}~\bibnamefont {Langacker}},\
  }\href@noop {} {\  (\bibinfo {year} {2015})},\ \Eprint
  {http://arxiv.org/abs/1512.07622} {arXiv:1512.07622 [hep-ph]} \BibitemShut
  {NoStop}%
\bibitem [{\citenamefont {Allanach}\ \emph {et~al.}(2015)\citenamefont
  {Allanach}, \citenamefont {Dev}, \citenamefont {Renner},\ and\ \citenamefont
  {Sakurai}}]{Allanach:2015ixl}%
  \BibitemOpen
  \bibfield  {author} {\bibinfo {author} {\bibfnamefont {B.~C.}\ \bibnamefont
  {Allanach}}, \bibinfo {author} {\bibfnamefont {P.~S.~B.}\ \bibnamefont
  {Dev}}, \bibinfo {author} {\bibfnamefont {S.~A.}\ \bibnamefont {Renner}}, \
  and\ \bibinfo {author} {\bibfnamefont {K.}~\bibnamefont {Sakurai}},\
  }\href@noop {} {\  (\bibinfo {year} {2015})},\ \Eprint
  {http://arxiv.org/abs/1512.07645} {arXiv:1512.07645 [hep-ph]} \BibitemShut
  {NoStop}%
\bibitem [{\citenamefont {Das}\ and\ \citenamefont {Rai}(2015)}]{Das:2015enc}%
  \BibitemOpen
  \bibfield  {author} {\bibinfo {author} {\bibfnamefont {K.}~\bibnamefont
  {Das}}\ and\ \bibinfo {author} {\bibfnamefont {S.~K.}\ \bibnamefont {Rai}},\
  }\href@noop {} {\  (\bibinfo {year} {2015})},\ \Eprint
  {http://arxiv.org/abs/1512.07789} {arXiv:1512.07789 [hep-ph]} \BibitemShut
  {NoStop}%
\bibitem [{\citenamefont {Cheung}\ \emph {et~al.}(2015)\citenamefont {Cheung},
  \citenamefont {Ko}, \citenamefont {Lee}, \citenamefont {Park},\ and\
  \citenamefont {Tseng}}]{Cheung:2015cug}%
  \BibitemOpen
  \bibfield  {author} {\bibinfo {author} {\bibfnamefont {K.}~\bibnamefont
  {Cheung}}, \bibinfo {author} {\bibfnamefont {P.}~\bibnamefont {Ko}}, \bibinfo
  {author} {\bibfnamefont {J.~S.}\ \bibnamefont {Lee}}, \bibinfo {author}
  {\bibfnamefont {J.}~\bibnamefont {Park}}, \ and\ \bibinfo {author}
  {\bibfnamefont {P.-Y.}\ \bibnamefont {Tseng}},\ }\href@noop {} {\  (\bibinfo
  {year} {2015})},\ \Eprint {http://arxiv.org/abs/1512.07853} {arXiv:1512.07853
  [hep-ph]} \BibitemShut {NoStop}%
\bibitem [{\citenamefont {Liu}\ \emph {et~al.}(2015)\citenamefont {Liu},
  \citenamefont {Wang},\ and\ \citenamefont {Xue}}]{Liu:2015yec}%
  \BibitemOpen
  \bibfield  {author} {\bibinfo {author} {\bibfnamefont {J.}~\bibnamefont
  {Liu}}, \bibinfo {author} {\bibfnamefont {X.-P.}\ \bibnamefont {Wang}}, \
  and\ \bibinfo {author} {\bibfnamefont {W.}~\bibnamefont {Xue}},\ }\href@noop
  {} {\  (\bibinfo {year} {2015})},\ \Eprint {http://arxiv.org/abs/1512.07885}
  {arXiv:1512.07885 [hep-ph]} \BibitemShut {NoStop}%
\bibitem [{\citenamefont {Zhang}\ and\ \citenamefont
  {Zhou}(2015)}]{Zhang:2015uuo}%
  \BibitemOpen
  \bibfield  {author} {\bibinfo {author} {\bibfnamefont {J.}~\bibnamefont
  {Zhang}}\ and\ \bibinfo {author} {\bibfnamefont {S.}~\bibnamefont {Zhou}},\
  }\href@noop {} {\  (\bibinfo {year} {2015})},\ \Eprint
  {http://arxiv.org/abs/1512.07889} {arXiv:1512.07889 [hep-ph]} \BibitemShut
  {NoStop}%
\bibitem [{\citenamefont {An}\ \emph {et~al.}(2015)\citenamefont {An},
  \citenamefont {Cheung},\ and\ \citenamefont {Zhang}}]{An:2015cgp}%
  \BibitemOpen
  \bibfield  {author} {\bibinfo {author} {\bibfnamefont {H.}~\bibnamefont
  {An}}, \bibinfo {author} {\bibfnamefont {C.}~\bibnamefont {Cheung}}, \ and\
  \bibinfo {author} {\bibfnamefont {Y.}~\bibnamefont {Zhang}},\ }\href@noop {}
  {\  (\bibinfo {year} {2015})},\ \Eprint {http://arxiv.org/abs/1512.08378}
  {arXiv:1512.08378 [hep-ph]} \BibitemShut {NoStop}%
\bibitem [{\citenamefont {Wang}\ \emph
  {et~al.}(2015{\natexlab{b}})\citenamefont {Wang}, \citenamefont {Wang},
  \citenamefont {Wu}, \citenamefont {Yang},\ and\ \citenamefont
  {Zhang}}]{Wang:2015omi}%
  \BibitemOpen
  \bibfield  {author} {\bibinfo {author} {\bibfnamefont {F.}~\bibnamefont
  {Wang}}, \bibinfo {author} {\bibfnamefont {W.}~\bibnamefont {Wang}}, \bibinfo
  {author} {\bibfnamefont {L.}~\bibnamefont {Wu}}, \bibinfo {author}
  {\bibfnamefont {J.~M.}\ \bibnamefont {Yang}}, \ and\ \bibinfo {author}
  {\bibfnamefont {M.}~\bibnamefont {Zhang}},\ }\href@noop {} {\  (\bibinfo
  {year} {2015}{\natexlab{b}})},\ \Eprint {http://arxiv.org/abs/1512.08434}
  {arXiv:1512.08434 [hep-ph]} \BibitemShut {NoStop}%
\bibitem [{\citenamefont {Cao}\ \emph {et~al.}(2015{\natexlab{d}})\citenamefont
  {Cao}, \citenamefont {Liu}, \citenamefont {Xie}, \citenamefont {Yan},\ and\
  \citenamefont {Zhang}}]{Cao:2015scs}%
  \BibitemOpen
  \bibfield  {author} {\bibinfo {author} {\bibfnamefont {Q.-H.}\ \bibnamefont
  {Cao}}, \bibinfo {author} {\bibfnamefont {Y.}~\bibnamefont {Liu}}, \bibinfo
  {author} {\bibfnamefont {K.-P.}\ \bibnamefont {Xie}}, \bibinfo {author}
  {\bibfnamefont {B.}~\bibnamefont {Yan}}, \ and\ \bibinfo {author}
  {\bibfnamefont {D.-M.}\ \bibnamefont {Zhang}},\ }\href@noop {} {\  (\bibinfo
  {year} {2015}{\natexlab{d}})},\ \Eprint {http://arxiv.org/abs/1512.08441}
  {arXiv:1512.08441 [hep-ph]} \BibitemShut {NoStop}%
\bibitem [{\citenamefont {Gao}\ \emph {et~al.}(2015)\citenamefont {Gao},
  \citenamefont {Zhang},\ and\ \citenamefont {Zhu}}]{Gao:2015igz}%
  \BibitemOpen
  \bibfield  {author} {\bibinfo {author} {\bibfnamefont {J.}~\bibnamefont
  {Gao}}, \bibinfo {author} {\bibfnamefont {H.}~\bibnamefont {Zhang}}, \ and\
  \bibinfo {author} {\bibfnamefont {H.~X.}\ \bibnamefont {Zhu}},\ }\href@noop
  {} {\  (\bibinfo {year} {2015})},\ \Eprint {http://arxiv.org/abs/1512.08478}
  {arXiv:1512.08478 [hep-ph]} \BibitemShut {NoStop}%
\bibitem [{\citenamefont {Goertz}\ \emph {et~al.}(2015)\citenamefont {Goertz},
  \citenamefont {Kamenik}, \citenamefont {Katz},\ and\ \citenamefont
  {Nardecchia}}]{Goertz:2015nkp}%
  \BibitemOpen
  \bibfield  {author} {\bibinfo {author} {\bibfnamefont {F.}~\bibnamefont
  {Goertz}}, \bibinfo {author} {\bibfnamefont {J.~F.}\ \bibnamefont {Kamenik}},
  \bibinfo {author} {\bibfnamefont {A.}~\bibnamefont {Katz}}, \ and\ \bibinfo
  {author} {\bibfnamefont {M.}~\bibnamefont {Nardecchia}},\ }\href@noop {} {\
  (\bibinfo {year} {2015})},\ \Eprint {http://arxiv.org/abs/1512.08500}
  {arXiv:1512.08500 [hep-ph]} \BibitemShut {NoStop}%
\bibitem [{\citenamefont {Dev}\ \emph {et~al.}(2015)\citenamefont {Dev},
  \citenamefont {Mohapatra},\ and\ \citenamefont {Zhang}}]{Dev:2015vjd}%
  \BibitemOpen
  \bibfield  {author} {\bibinfo {author} {\bibfnamefont {P.~S.~B.}\
  \bibnamefont {Dev}}, \bibinfo {author} {\bibfnamefont {R.~N.}\ \bibnamefont
  {Mohapatra}}, \ and\ \bibinfo {author} {\bibfnamefont {Y.}~\bibnamefont
  {Zhang}},\ }\href@noop {} {\  (\bibinfo {year} {2015})},\ \Eprint
  {http://arxiv.org/abs/1512.08507} {arXiv:1512.08507 [hep-ph]} \BibitemShut
  {NoStop}%
\bibitem [{\citenamefont {Li}\ \emph {et~al.}(2015)\citenamefont {Li},
  \citenamefont {Mao}, \citenamefont {Tang}, \citenamefont {Zhang},
  \citenamefont {Zhou},\ and\ \citenamefont {Zhu}}]{Li:2015jwd}%
  \BibitemOpen
  \bibfield  {author} {\bibinfo {author} {\bibfnamefont {G.}~\bibnamefont
  {Li}}, \bibinfo {author} {\bibfnamefont {Y.-n.}\ \bibnamefont {Mao}},
  \bibinfo {author} {\bibfnamefont {Y.-L.}\ \bibnamefont {Tang}}, \bibinfo
  {author} {\bibfnamefont {C.}~\bibnamefont {Zhang}}, \bibinfo {author}
  {\bibfnamefont {Y.}~\bibnamefont {Zhou}}, \ and\ \bibinfo {author}
  {\bibfnamefont {S.-h.}\ \bibnamefont {Zhu}},\ }\href@noop {} {\  (\bibinfo
  {year} {2015})},\ \Eprint {http://arxiv.org/abs/1512.08255} {arXiv:1512.08255
  [hep-ph]} \BibitemShut {NoStop}%
\bibitem [{\citenamefont {Son}\ and\ \citenamefont
  {Urbano}(2015)}]{Son:2015vfl}%
  \BibitemOpen
  \bibfield  {author} {\bibinfo {author} {\bibfnamefont {M.}~\bibnamefont
  {Son}}\ and\ \bibinfo {author} {\bibfnamefont {A.}~\bibnamefont {Urbano}},\
  }\href@noop {} {\  (\bibinfo {year} {2015})},\ \Eprint
  {http://arxiv.org/abs/1512.08307} {arXiv:1512.08307 [hep-ph]} \BibitemShut
  {NoStop}%
\bibitem [{\citenamefont {Tang}\ and\ \citenamefont
  {Zhu}(2015)}]{Tang:2015eko}%
  \BibitemOpen
  \bibfield  {author} {\bibinfo {author} {\bibfnamefont {Y.-L.}\ \bibnamefont
  {Tang}}\ and\ \bibinfo {author} {\bibfnamefont {S.-h.}\ \bibnamefont {Zhu}},\
  }\href@noop {} {\  (\bibinfo {year} {2015})},\ \Eprint
  {http://arxiv.org/abs/1512.08323} {arXiv:1512.08323 [hep-ph]} \BibitemShut
  {NoStop}%
\bibitem [{\citenamefont {Cao}\ \emph {et~al.}(2015{\natexlab{e}})\citenamefont
  {Cao}, \citenamefont {Wang},\ and\ \citenamefont {Zhang}}]{Cao:2015apa}%
  \BibitemOpen
  \bibfield  {author} {\bibinfo {author} {\bibfnamefont {J.}~\bibnamefont
  {Cao}}, \bibinfo {author} {\bibfnamefont {F.}~\bibnamefont {Wang}}, \ and\
  \bibinfo {author} {\bibfnamefont {Y.}~\bibnamefont {Zhang}},\ }\href@noop {}
  {\  (\bibinfo {year} {2015}{\natexlab{e}})},\ \Eprint
  {http://arxiv.org/abs/1512.08392} {arXiv:1512.08392 [hep-ph]} \BibitemShut
  {NoStop}%
\bibitem [{\citenamefont {Cai}\ \emph {et~al.}(2015)\citenamefont {Cai},
  \citenamefont {Yu},\ and\ \citenamefont {Zhang}}]{Cai:2015hzc}%
  \BibitemOpen
  \bibfield  {author} {\bibinfo {author} {\bibfnamefont {C.}~\bibnamefont
  {Cai}}, \bibinfo {author} {\bibfnamefont {Z.-H.}\ \bibnamefont {Yu}}, \ and\
  \bibinfo {author} {\bibfnamefont {H.-H.}\ \bibnamefont {Zhang}},\ }\href@noop
  {} {\  (\bibinfo {year} {2015})},\ \Eprint {http://arxiv.org/abs/1512.08440}
  {arXiv:1512.08440 [hep-ph]} \BibitemShut {NoStop}%
\bibitem [{\citenamefont {Chao}(2015{\natexlab{b}})}]{Chao:2015nac}%
  \BibitemOpen
  \bibfield  {author} {\bibinfo {author} {\bibfnamefont {W.}~\bibnamefont
  {Chao}},\ }\href@noop {} {\  (\bibinfo {year} {2015}{\natexlab{b}})},\
  \Eprint {http://arxiv.org/abs/1512.08484} {arXiv:1512.08484 [hep-ph]}
  \BibitemShut {NoStop}%
\bibitem [{\citenamefont {Anchordoqui}\ \emph {et~al.}(2015)\citenamefont
  {Anchordoqui}, \citenamefont {Antoniadis}, \citenamefont {Goldberg},
  \citenamefont {Huang}, \citenamefont {Lust},\ and\ \citenamefont
  {Taylor}}]{Anchordoqui:2015jxc}%
  \BibitemOpen
  \bibfield  {author} {\bibinfo {author} {\bibfnamefont {L.~A.}\ \bibnamefont
  {Anchordoqui}}, \bibinfo {author} {\bibfnamefont {I.}~\bibnamefont
  {Antoniadis}}, \bibinfo {author} {\bibfnamefont {H.}~\bibnamefont
  {Goldberg}}, \bibinfo {author} {\bibfnamefont {X.}~\bibnamefont {Huang}},
  \bibinfo {author} {\bibfnamefont {D.}~\bibnamefont {Lust}}, \ and\ \bibinfo
  {author} {\bibfnamefont {T.~R.}\ \bibnamefont {Taylor}},\ }\href@noop {} {\
  (\bibinfo {year} {2015})},\ \Eprint {http://arxiv.org/abs/1512.08502}
  {arXiv:1512.08502 [hep-ph]} \BibitemShut {NoStop}%
\bibitem [{\citenamefont {Bizot}\ \emph {et~al.}(2015)\citenamefont {Bizot},
  \citenamefont {Davidson}, \citenamefont {Frigerio},\ and\ \citenamefont
  {Kneur}}]{Bizot:2015qqo}%
  \BibitemOpen
  \bibfield  {author} {\bibinfo {author} {\bibfnamefont {N.}~\bibnamefont
  {Bizot}}, \bibinfo {author} {\bibfnamefont {S.}~\bibnamefont {Davidson}},
  \bibinfo {author} {\bibfnamefont {M.}~\bibnamefont {Frigerio}}, \ and\
  \bibinfo {author} {\bibfnamefont {J.~L.}\ \bibnamefont {Kneur}},\ }\href@noop
  {} {\  (\bibinfo {year} {2015})},\ \Eprint {http://arxiv.org/abs/1512.08508}
  {arXiv:1512.08508 [hep-ph]} \BibitemShut {NoStop}%
\bibitem [{\citenamefont {Ibanez}\ and\ \citenamefont
  {Martin-Lozano}(2015)}]{Ibanez:2015uok}%
  \BibitemOpen
  \bibfield  {author} {\bibinfo {author} {\bibfnamefont {L.~E.}\ \bibnamefont
  {Ibanez}}\ and\ \bibinfo {author} {\bibfnamefont {V.}~\bibnamefont
  {Martin-Lozano}},\ }\href@noop {} {\  (\bibinfo {year} {2015})},\ \Eprint
  {http://arxiv.org/abs/1512.08777} {arXiv:1512.08777 [hep-ph]} \BibitemShut
  {NoStop}%
\bibitem [{\citenamefont {Kang}\ and\ \citenamefont
  {Song}(2015)}]{Kang:2015roj}%
  \BibitemOpen
  \bibfield  {author} {\bibinfo {author} {\bibfnamefont {S.~K.}\ \bibnamefont
  {Kang}}\ and\ \bibinfo {author} {\bibfnamefont {J.}~\bibnamefont {Song}},\
  }\href@noop {} {\  (\bibinfo {year} {2015})},\ \Eprint
  {http://arxiv.org/abs/1512.08963} {arXiv:1512.08963 [hep-ph]} \BibitemShut
  {NoStop}%
\bibitem [{\citenamefont {Kanemura}\ \emph
  {et~al.}(2015{\natexlab{a}})\citenamefont {Kanemura}, \citenamefont
  {Nishiwaki}, \citenamefont {Okada}, \citenamefont {Orikasa}, \citenamefont
  {Park},\ and\ \citenamefont {Watanabe}}]{Kanemura:2015bli}%
  \BibitemOpen
  \bibfield  {author} {\bibinfo {author} {\bibfnamefont {S.}~\bibnamefont
  {Kanemura}}, \bibinfo {author} {\bibfnamefont {K.}~\bibnamefont {Nishiwaki}},
  \bibinfo {author} {\bibfnamefont {H.}~\bibnamefont {Okada}}, \bibinfo
  {author} {\bibfnamefont {Y.}~\bibnamefont {Orikasa}}, \bibinfo {author}
  {\bibfnamefont {S.~C.}\ \bibnamefont {Park}}, \ and\ \bibinfo {author}
  {\bibfnamefont {R.}~\bibnamefont {Watanabe}},\ }\href@noop {} {\  (\bibinfo
  {year} {2015}{\natexlab{a}})},\ \Eprint {http://arxiv.org/abs/1512.09048}
  {arXiv:1512.09048 [hep-ph]} \BibitemShut {NoStop}%
\bibitem [{\citenamefont {Low}\ and\ \citenamefont
  {Lykken}(2015)}]{Low:2015qho}%
  \BibitemOpen
  \bibfield  {author} {\bibinfo {author} {\bibfnamefont {I.}~\bibnamefont
  {Low}}\ and\ \bibinfo {author} {\bibfnamefont {J.}~\bibnamefont {Lykken}},\
  }\href@noop {} {\  (\bibinfo {year} {2015})},\ \Eprint
  {http://arxiv.org/abs/1512.09089} {arXiv:1512.09089 [hep-ph]} \BibitemShut
  {NoStop}%
\bibitem [{\citenamefont {Kaneta}\ \emph {et~al.}(2015)\citenamefont {Kaneta},
  \citenamefont {Kang},\ and\ \citenamefont {Lee}}]{Kaneta:2015qpf}%
  \BibitemOpen
  \bibfield  {author} {\bibinfo {author} {\bibfnamefont {K.}~\bibnamefont
  {Kaneta}}, \bibinfo {author} {\bibfnamefont {S.}~\bibnamefont {Kang}}, \ and\
  \bibinfo {author} {\bibfnamefont {H.-S.}\ \bibnamefont {Lee}},\ }\href@noop
  {} {\  (\bibinfo {year} {2015})},\ \Eprint {http://arxiv.org/abs/1512.09129}
  {arXiv:1512.09129 [hep-ph]} \BibitemShut {NoStop}%
\bibitem [{\citenamefont {Dasgupta}\ \emph {et~al.}(2015)\citenamefont
  {Dasgupta}, \citenamefont {Mitra},\ and\ \citenamefont
  {Borah}}]{Dasgupta:2015pbr}%
  \BibitemOpen
  \bibfield  {author} {\bibinfo {author} {\bibfnamefont {A.}~\bibnamefont
  {Dasgupta}}, \bibinfo {author} {\bibfnamefont {M.}~\bibnamefont {Mitra}}, \
  and\ \bibinfo {author} {\bibfnamefont {D.}~\bibnamefont {Borah}},\
  }\href@noop {} {\  (\bibinfo {year} {2015})},\ \Eprint
  {http://arxiv.org/abs/1512.09202} {arXiv:1512.09202 [hep-ph]} \BibitemShut
  {NoStop}%
\bibitem [{\citenamefont {Jung}\ \emph {et~al.}(2015)\citenamefont {Jung},
  \citenamefont {Song},\ and\ \citenamefont {Yoon}}]{Jung:2015etr}%
  \BibitemOpen
  \bibfield  {author} {\bibinfo {author} {\bibfnamefont {S.}~\bibnamefont
  {Jung}}, \bibinfo {author} {\bibfnamefont {J.}~\bibnamefont {Song}}, \ and\
  \bibinfo {author} {\bibfnamefont {Y.~W.}\ \bibnamefont {Yoon}},\ }\href@noop
  {} {\  (\bibinfo {year} {2015})},\ \Eprint {http://arxiv.org/abs/1601.00006}
  {arXiv:1601.00006 [hep-ph]} \BibitemShut {NoStop}%
\bibitem [{\citenamefont {Nomura}\ and\ \citenamefont
  {Okada}(2016)}]{Nomura:2016fzs}%
  \BibitemOpen
  \bibfield  {author} {\bibinfo {author} {\bibfnamefont {T.}~\bibnamefont
  {Nomura}}\ and\ \bibinfo {author} {\bibfnamefont {H.}~\bibnamefont {Okada}},\
  }\href@noop {} {\  (\bibinfo {year} {2016})},\ \Eprint
  {http://arxiv.org/abs/1601.00386} {arXiv:1601.00386 [hep-ph]} \BibitemShut
  {NoStop}%
\bibitem [{\citenamefont {Ko}\ \emph {et~al.}(2016)\citenamefont {Ko},
  \citenamefont {Omura},\ and\ \citenamefont {Yu}}]{Ko:2016lai}%
  \BibitemOpen
  \bibfield  {author} {\bibinfo {author} {\bibfnamefont {P.}~\bibnamefont
  {Ko}}, \bibinfo {author} {\bibfnamefont {Y.}~\bibnamefont {Omura}}, \ and\
  \bibinfo {author} {\bibfnamefont {C.}~\bibnamefont {Yu}},\ }\href@noop {} {\
  (\bibinfo {year} {2016})},\ \Eprint {http://arxiv.org/abs/1601.00586}
  {arXiv:1601.00586 [hep-ph]} \BibitemShut {NoStop}%
\bibitem [{\citenamefont {Palti}(2016)}]{Palti:2016kew}%
  \BibitemOpen
  \bibfield  {author} {\bibinfo {author} {\bibfnamefont {E.}~\bibnamefont
  {Palti}},\ }\href@noop {} {\  (\bibinfo {year} {2016})},\ \Eprint
  {http://arxiv.org/abs/1601.00285} {arXiv:1601.00285 [hep-ph]} \BibitemShut
  {NoStop}%
\bibitem [{\citenamefont {Han}\ \emph {et~al.}(2016)\citenamefont {Han},
  \citenamefont {Wang}, \citenamefont {Wu}, \citenamefont {Yang},\ and\
  \citenamefont {Zhang}}]{Han:2016bus}%
  \BibitemOpen
  \bibfield  {author} {\bibinfo {author} {\bibfnamefont {X.-F.}\ \bibnamefont
  {Han}}, \bibinfo {author} {\bibfnamefont {L.}~\bibnamefont {Wang}}, \bibinfo
  {author} {\bibfnamefont {L.}~\bibnamefont {Wu}}, \bibinfo {author}
  {\bibfnamefont {J.~M.}\ \bibnamefont {Yang}}, \ and\ \bibinfo {author}
  {\bibfnamefont {M.}~\bibnamefont {Zhang}},\ }\href@noop {} {\  (\bibinfo
  {year} {2016})},\ \Eprint {http://arxiv.org/abs/1601.00534} {arXiv:1601.00534
  [hep-ph]} \BibitemShut {NoStop}%
\bibitem [{\citenamefont {Danielsson}\ \emph {et~al.}(2016)\citenamefont
  {Danielsson}, \citenamefont {Enberg}, \citenamefont {Ingelman},\ and\
  \citenamefont {Mandal}}]{Danielsson:2016nyy}%
  \BibitemOpen
  \bibfield  {author} {\bibinfo {author} {\bibfnamefont {U.}~\bibnamefont
  {Danielsson}}, \bibinfo {author} {\bibfnamefont {R.}~\bibnamefont {Enberg}},
  \bibinfo {author} {\bibfnamefont {G.}~\bibnamefont {Ingelman}}, \ and\
  \bibinfo {author} {\bibfnamefont {T.}~\bibnamefont {Mandal}},\ }\href@noop {}
  {\  (\bibinfo {year} {2016})},\ \Eprint {http://arxiv.org/abs/1601.00624}
  {arXiv:1601.00624 [hep-ph]} \BibitemShut {NoStop}%
\bibitem [{\citenamefont {Chao}(2016)}]{Chao:2016mtn}%
  \BibitemOpen
  \bibfield  {author} {\bibinfo {author} {\bibfnamefont {W.}~\bibnamefont
  {Chao}},\ }\href@noop {} {\  (\bibinfo {year} {2016})},\ \Eprint
  {http://arxiv.org/abs/1601.00633} {arXiv:1601.00633 [hep-ph]} \BibitemShut
  {NoStop}%
\bibitem [{\citenamefont {Csaki}\ \emph {et~al.}(2016)\citenamefont {Csaki},
  \citenamefont {Hubisz}, \citenamefont {Lombardo},\ and\ \citenamefont
  {Terning}}]{Csaki:2016raa}%
  \BibitemOpen
  \bibfield  {author} {\bibinfo {author} {\bibfnamefont {C.}~\bibnamefont
  {Csaki}}, \bibinfo {author} {\bibfnamefont {J.}~\bibnamefont {Hubisz}},
  \bibinfo {author} {\bibfnamefont {S.}~\bibnamefont {Lombardo}}, \ and\
  \bibinfo {author} {\bibfnamefont {J.}~\bibnamefont {Terning}},\ }\href@noop
  {} {\  (\bibinfo {year} {2016})},\ \Eprint {http://arxiv.org/abs/1601.00638}
  {arXiv:1601.00638 [hep-ph]} \BibitemShut {NoStop}%
\bibitem [{\citenamefont {Karozas}\ \emph {et~al.}(2016)\citenamefont
  {Karozas}, \citenamefont {King}, \citenamefont {Leontaris},\ and\
  \citenamefont {Meadowcroft}}]{Karozas:2016hcp}%
  \BibitemOpen
  \bibfield  {author} {\bibinfo {author} {\bibfnamefont {A.}~\bibnamefont
  {Karozas}}, \bibinfo {author} {\bibfnamefont {S.~F.}\ \bibnamefont {King}},
  \bibinfo {author} {\bibfnamefont {G.~K.}\ \bibnamefont {Leontaris}}, \ and\
  \bibinfo {author} {\bibfnamefont {A.~K.}\ \bibnamefont {Meadowcroft}},\
  }\href@noop {} {\  (\bibinfo {year} {2016})},\ \Eprint
  {http://arxiv.org/abs/1601.00640} {arXiv:1601.00640 [hep-ph]} \BibitemShut
  {NoStop}%
\bibitem [{\citenamefont {Dutta}\ \emph {et~al.}(2016)\citenamefont {Dutta},
  \citenamefont {Gao}, \citenamefont {Ghosh}, \citenamefont {Gogoladze},
  \citenamefont {Li}, \citenamefont {Shafi},\ and\ \citenamefont
  {Walker}}]{Dutta:2016jqn}%
  \BibitemOpen
  \bibfield  {author} {\bibinfo {author} {\bibfnamefont {B.}~\bibnamefont
  {Dutta}}, \bibinfo {author} {\bibfnamefont {Y.}~\bibnamefont {Gao}}, \bibinfo
  {author} {\bibfnamefont {T.}~\bibnamefont {Ghosh}}, \bibinfo {author}
  {\bibfnamefont {I.}~\bibnamefont {Gogoladze}}, \bibinfo {author}
  {\bibfnamefont {T.}~\bibnamefont {Li}}, \bibinfo {author} {\bibfnamefont
  {Q.}~\bibnamefont {Shafi}}, \ and\ \bibinfo {author} {\bibfnamefont {J.~W.}\
  \bibnamefont {Walker}},\ }\href@noop {} {\  (\bibinfo {year} {2016})},\
  \Eprint {http://arxiv.org/abs/1601.00866} {arXiv:1601.00866 [hep-ph]}
  \BibitemShut {NoStop}%
\bibitem [{\citenamefont {Deppisch}\ \emph {et~al.}(2016)\citenamefont
  {Deppisch}, \citenamefont {Hati}, \citenamefont {Patra}, \citenamefont
  {Pritimita},\ and\ \citenamefont {Sarkar}}]{Deppisch:2016scs}%
  \BibitemOpen
  \bibfield  {author} {\bibinfo {author} {\bibfnamefont {F.~F.}\ \bibnamefont
  {Deppisch}}, \bibinfo {author} {\bibfnamefont {C.}~\bibnamefont {Hati}},
  \bibinfo {author} {\bibfnamefont {S.}~\bibnamefont {Patra}}, \bibinfo
  {author} {\bibfnamefont {P.}~\bibnamefont {Pritimita}}, \ and\ \bibinfo
  {author} {\bibfnamefont {U.}~\bibnamefont {Sarkar}},\ }\href@noop {} {\
  (\bibinfo {year} {2016})},\ \Eprint {http://arxiv.org/abs/1601.00952}
  {arXiv:1601.00952 [hep-ph]} \BibitemShut {NoStop}%
\bibitem [{\citenamefont {Ito}\ \emph {et~al.}(2016)\citenamefont {Ito},
  \citenamefont {Moroi},\ and\ \citenamefont {Takaesu}}]{Ito:2016zkz}%
  \BibitemOpen
  \bibfield  {author} {\bibinfo {author} {\bibfnamefont {H.}~\bibnamefont
  {Ito}}, \bibinfo {author} {\bibfnamefont {T.}~\bibnamefont {Moroi}}, \ and\
  \bibinfo {author} {\bibfnamefont {Y.}~\bibnamefont {Takaesu}},\ }\href@noop
  {} {\  (\bibinfo {year} {2016})},\ \Eprint {http://arxiv.org/abs/1601.01144}
  {arXiv:1601.01144 [hep-ph]} \BibitemShut {NoStop}%
\bibitem [{\citenamefont {Kanemura}\ \emph
  {et~al.}(2015{\natexlab{b}})\citenamefont {Kanemura}, \citenamefont
  {Machida}, \citenamefont {Odori},\ and\ \citenamefont
  {Shindou}}]{Kanemura:2015vcb}%
  \BibitemOpen
  \bibfield  {author} {\bibinfo {author} {\bibfnamefont {S.}~\bibnamefont
  {Kanemura}}, \bibinfo {author} {\bibfnamefont {N.}~\bibnamefont {Machida}},
  \bibinfo {author} {\bibfnamefont {S.}~\bibnamefont {Odori}}, \ and\ \bibinfo
  {author} {\bibfnamefont {T.}~\bibnamefont {Shindou}},\ }\href@noop {} {\
  (\bibinfo {year} {2015}{\natexlab{b}})},\ \Eprint
  {http://arxiv.org/abs/1512.09053} {arXiv:1512.09053 [hep-ph]} \BibitemShut
  {NoStop}%
\bibitem [{\citenamefont {Bi}\ \emph {et~al.}(2015{\natexlab{b}})\citenamefont
  {Bi}, \citenamefont {Ding}, \citenamefont {Fan}, \citenamefont {Huang},
  \citenamefont {Li}, \citenamefont {Li}, \citenamefont {Raza}, \citenamefont
  {Wang},\ and\ \citenamefont {Zhu}}]{Bi:2015lcf}%
  \BibitemOpen
  \bibfield  {author} {\bibinfo {author} {\bibfnamefont {X.-J.}\ \bibnamefont
  {Bi}}, \bibinfo {author} {\bibfnamefont {R.}~\bibnamefont {Ding}}, \bibinfo
  {author} {\bibfnamefont {Y.}~\bibnamefont {Fan}}, \bibinfo {author}
  {\bibfnamefont {L.}~\bibnamefont {Huang}}, \bibinfo {author} {\bibfnamefont
  {C.}~\bibnamefont {Li}}, \bibinfo {author} {\bibfnamefont {T.}~\bibnamefont
  {Li}}, \bibinfo {author} {\bibfnamefont {S.}~\bibnamefont {Raza}}, \bibinfo
  {author} {\bibfnamefont {X.-C.}\ \bibnamefont {Wang}}, \ and\ \bibinfo
  {author} {\bibfnamefont {B.}~\bibnamefont {Zhu}},\ }\href@noop {} {\
  (\bibinfo {year} {2015}{\natexlab{b}})},\ \Eprint
  {http://arxiv.org/abs/1512.08497} {arXiv:1512.08497 [hep-ph]} \BibitemShut
  {NoStop}%
\bibitem [{\citenamefont {Antipin}\ \emph {et~al.}(2015)\citenamefont
  {Antipin}, \citenamefont {Mojaza},\ and\ \citenamefont
  {Sannino}}]{Antipin:2015kgh}%
  \BibitemOpen
  \bibfield  {author} {\bibinfo {author} {\bibfnamefont {O.}~\bibnamefont
  {Antipin}}, \bibinfo {author} {\bibfnamefont {M.}~\bibnamefont {Mojaza}}, \
  and\ \bibinfo {author} {\bibfnamefont {F.}~\bibnamefont {Sannino}},\
  }\href@noop {} {\  (\bibinfo {year} {2015})},\ \Eprint
  {http://arxiv.org/abs/1512.06708} {arXiv:1512.06708 [hep-ph]} \BibitemShut
  {NoStop}%
\bibitem [{\citenamefont {Hamada}\ \emph {et~al.}(2015)\citenamefont {Hamada},
  \citenamefont {Noumi}, \citenamefont {Sun},\ and\ \citenamefont
  {Shiu}}]{Hamada:2015skp}%
  \BibitemOpen
  \bibfield  {author} {\bibinfo {author} {\bibfnamefont {Y.}~\bibnamefont
  {Hamada}}, \bibinfo {author} {\bibfnamefont {T.}~\bibnamefont {Noumi}},
  \bibinfo {author} {\bibfnamefont {S.}~\bibnamefont {Sun}}, \ and\ \bibinfo
  {author} {\bibfnamefont {G.}~\bibnamefont {Shiu}},\ }\href@noop {} {\
  (\bibinfo {year} {2015})},\ \Eprint {http://arxiv.org/abs/1512.08984}
  {arXiv:1512.08984 [hep-ph]} \BibitemShut {NoStop}%
\bibitem [{\citenamefont {Dev}\ and\ \citenamefont
  {Teresi}(2015)}]{Dev:2015isx}%
  \BibitemOpen
  \bibfield  {author} {\bibinfo {author} {\bibfnamefont {P.~S.~B.}\
  \bibnamefont {Dev}}\ and\ \bibinfo {author} {\bibfnamefont {D.}~\bibnamefont
  {Teresi}},\ }\href@noop {} {\  (\bibinfo {year} {2015})},\ \Eprint
  {http://arxiv.org/abs/1512.07243} {arXiv:1512.07243 [hep-ph]} \BibitemShut
  {NoStop}%
\bibitem [{\citenamefont {Hernández}\ \emph {et~al.}(2016)\citenamefont
  {Hernández}, \citenamefont {Varzielas},\ and\ \citenamefont
  {Schumacher}}]{Hernandez:2016rbi}%
  \BibitemOpen
  \bibfield  {author} {\bibinfo {author} {\bibfnamefont {A.~E.~C.}\
  \bibnamefont {Hernández}}, \bibinfo {author} {\bibfnamefont {I.~d.~M.}\
  \bibnamefont {Varzielas}}, \ and\ \bibinfo {author} {\bibfnamefont
  {E.}~\bibnamefont {Schumacher}},\ }\href@noop {} {\  (\bibinfo {year}
  {2016})},\ \Eprint {http://arxiv.org/abs/1601.00661} {arXiv:1601.00661
  [hep-ph]} \BibitemShut {NoStop}%
\bibitem [{\citenamefont {Ghorbani}\ and\ \citenamefont
  {Ghorbani}(2016)}]{Ghorbani:2016jdq}%
  \BibitemOpen
  \bibfield  {author} {\bibinfo {author} {\bibfnamefont {K.}~\bibnamefont
  {Ghorbani}}\ and\ \bibinfo {author} {\bibfnamefont {H.}~\bibnamefont
  {Ghorbani}},\ }\href@noop {} {\  (\bibinfo {year} {2016})},\ \Eprint
  {http://arxiv.org/abs/1601.00602} {arXiv:1601.00602 [hep-ph]} \BibitemShut
  {NoStop}%
\bibitem [{\citenamefont {Modak}\ \emph {et~al.}(2016)\citenamefont {Modak},
  \citenamefont {Sadhukhan},\ and\ \citenamefont {Srivastava}}]{Modak:2016ung}%
  \BibitemOpen
  \bibfield  {author} {\bibinfo {author} {\bibfnamefont {T.}~\bibnamefont
  {Modak}}, \bibinfo {author} {\bibfnamefont {S.}~\bibnamefont {Sadhukhan}}, \
  and\ \bibinfo {author} {\bibfnamefont {R.}~\bibnamefont {Srivastava}},\
  }\href {\doibase 10.1016/j.physletb.2016.03.021} {\bibfield  {journal}
  {\bibinfo  {journal} {Phys. Lett.}\ }\textbf {\bibinfo {volume} {B756}},\
  \bibinfo {pages} {405} (\bibinfo {year} {2016})},\ \Eprint
  {http://arxiv.org/abs/1601.00836} {arXiv:1601.00836 [hep-ph]} \BibitemShut
  {NoStop}%
\bibitem [{\citenamefont {Fichet}\ \emph
  {et~al.}(2015{\natexlab{b}})\citenamefont {Fichet}, \citenamefont {von
  Gersdorff}, \citenamefont {Lenzi}, \citenamefont {Royon},\ and\ \citenamefont
  {Saimpert}}]{Fichet:2014uka}%
  \BibitemOpen
  \bibfield  {author} {\bibinfo {author} {\bibfnamefont {S.}~\bibnamefont
  {Fichet}}, \bibinfo {author} {\bibfnamefont {G.}~\bibnamefont {von
  Gersdorff}}, \bibinfo {author} {\bibfnamefont {B.}~\bibnamefont {Lenzi}},
  \bibinfo {author} {\bibfnamefont {C.}~\bibnamefont {Royon}}, \ and\ \bibinfo
  {author} {\bibfnamefont {M.}~\bibnamefont {Saimpert}},\ }\href {\doibase
  10.1007/JHEP02(2015)165} {\bibfield  {journal} {\bibinfo  {journal} {JHEP}\
  }\textbf {\bibinfo {volume} {02}},\ \bibinfo {pages} {165} (\bibinfo {year}
  {2015}{\natexlab{b}})},\ \Eprint {http://arxiv.org/abs/1411.6629}
  {arXiv:1411.6629 [hep-ph]} \BibitemShut {NoStop}%
\bibitem [{\citenamefont {Fichet}(2015)}]{Fichet:2015yba}%
  \BibitemOpen
  \bibfield  {author} {\bibinfo {author} {\bibfnamefont {S.}~\bibnamefont
  {Fichet}},\ }in\ \href
  {https://inspirehep.net/record/1353402/files/arXiv:1503.05219.pdf} {\emph
  {\bibinfo {booktitle} {{New Trends in High-Energy Physics and QCD Natal, Rio
  Grande do Norte, Brazil, October 21-November 6, 2014}}}}\ (\bibinfo {year}
  {2015})\ \Eprint {http://arxiv.org/abs/1503.05219} {arXiv:1503.05219
  [hep-ph]} \BibitemShut {NoStop}%
\bibitem [{\citenamefont {Fichet}\ and\ \citenamefont {von
  Gersdorff}(2015)}]{Fichet:2015yia}%
  \BibitemOpen
  \bibfield  {author} {\bibinfo {author} {\bibfnamefont {S.}~\bibnamefont
  {Fichet}}\ and\ \bibinfo {author} {\bibfnamefont {G.}~\bibnamefont {von
  Gersdorff}},\ }\href@noop {} {\  (\bibinfo {year} {2015})},\ \Eprint
  {http://arxiv.org/abs/1508.04814} {arXiv:1508.04814 [hep-ph]} \BibitemShut
  {NoStop}%
\bibitem [{Note1()}]{Note1}%
  \BibitemOpen
  \bibinfo {note} {For simplicity and to reduce the number of free parameters,
  we are assuming minimal flavour violation. This assumption is however not
  crucial for our present study.}\BibitemShut {Stop}%
\bibitem [{\citenamefont {Aad}\ \emph {et~al.}(2015)\citenamefont {Aad} \emph
  {et~al.}}]{ATLAS_dijet_8TeV}%
  \BibitemOpen
  \bibfield  {author} {\bibinfo {author} {\bibfnamefont {G.}~\bibnamefont
  {Aad}} \emph {et~al.} (\bibinfo {collaboration} {ATLAS}),\ }\href {\doibase
  10.1103/PhysRevD.91.052007} {\bibfield  {journal} {\bibinfo  {journal} {Phys.
  Rev.}\ }\textbf {\bibinfo {volume} {D91}},\ \bibinfo {pages} {052007}
  (\bibinfo {year} {2015})},\ \Eprint {http://arxiv.org/abs/1407.1376}
  {arXiv:1407.1376 [hep-ex]} \BibitemShut {NoStop}%
\bibitem [{CMS(2015{\natexlab{b}})}]{CMS_dijet_8TeV}%
  \BibitemOpen
  \href {http://cds.cern.ch/record/2063491} {\emph {\bibinfo {title} {{Search
  for Resonances Decaying to Dijet Final States at $\sqrt{s} = 8$ TeV with
  Scouting Data}}}},\ \bibinfo {type} {Tech. Rep.}\ \bibinfo {number}
  {CMS-PAS-EXO-14-005}\ (\bibinfo  {institution} {CERN},\ \bibinfo {address}
  {Geneva},\ \bibinfo {year} {2015})\BibitemShut {NoStop}%
\bibitem [{Note2()}]{Note2}%
  \BibitemOpen
  \bibinfo {note} {We will not assume the use of timing detectors for the
  simulations in this paper}\BibitemShut {NoStop}%
\bibitem [{\citenamefont {Boonekamp}\ \emph {et~al.}(2011)\citenamefont
  {Boonekamp}, \citenamefont {Dechambre}, \citenamefont {Juranek},
  \citenamefont {Kepka}, \citenamefont {Rangel} \emph {et~al.}}]{FPMC}%
  \BibitemOpen
  \bibfield  {author} {\bibinfo {author} {\bibfnamefont {M.}~\bibnamefont
  {Boonekamp}}, \bibinfo {author} {\bibfnamefont {A.}~\bibnamefont
  {Dechambre}}, \bibinfo {author} {\bibfnamefont {V.}~\bibnamefont {Juranek}},
  \bibinfo {author} {\bibfnamefont {O.}~\bibnamefont {Kepka}}, \bibinfo
  {author} {\bibfnamefont {M.}~\bibnamefont {Rangel}},  \emph {et~al.},\
  }\href@noop {} {\  (\bibinfo {year} {2011})},\ \Eprint
  {http://arxiv.org/abs/1102.2531} {arXiv:1102.2531 [hep-ph]} \BibitemShut
  {NoStop}%
\bibitem [{\citenamefont {Budnev}\ \emph {et~al.}(1975)\citenamefont {Budnev},
  \citenamefont {Ginzburg}, \citenamefont {Meledin},\ and\ \citenamefont
  {Serbo}}]{Budnev:1974de}%
  \BibitemOpen
  \bibfield  {author} {\bibinfo {author} {\bibfnamefont {V.~M.}\ \bibnamefont
  {Budnev}}, \bibinfo {author} {\bibfnamefont {I.~F.}\ \bibnamefont
  {Ginzburg}}, \bibinfo {author} {\bibfnamefont {G.~V.}\ \bibnamefont
  {Meledin}}, \ and\ \bibinfo {author} {\bibfnamefont {V.~G.}\ \bibnamefont
  {Serbo}},\ }\href {\doibase 10.1016/0370-1573(75)90009-5} {\bibfield
  {journal} {\bibinfo  {journal} {Phys. Rept.}\ }\textbf {\bibinfo {volume}
  {15}},\ \bibinfo {pages} {181} (\bibinfo {year} {1975})}\BibitemShut
  {NoStop}%
\bibitem [{\citenamefont {Khoze}\ \emph {et~al.}(2002)\citenamefont {Khoze},
  \citenamefont {Martin},\ and\ \citenamefont {Ryskin}}]{kmr}%
  \BibitemOpen
  \bibfield  {author} {\bibinfo {author} {\bibfnamefont {V.}~\bibnamefont
  {Khoze}}, \bibinfo {author} {\bibfnamefont {A.}~\bibnamefont {Martin}}, \
  and\ \bibinfo {author} {\bibfnamefont {M.}~\bibnamefont {Ryskin}},\ }\href
  {\doibase 10.1007/s100520100884} {\bibfield  {journal} {\bibinfo  {journal}
  {Eur.Phys.J.}\ }\textbf {\bibinfo {volume} {C23}},\ \bibinfo {pages} {311}
  (\bibinfo {year} {2002})},\ \Eprint {http://arxiv.org/abs/hep-ph/0111078}
  {arXiv:hep-ph/0111078 [hep-ph]} \BibitemShut {NoStop}%
\bibitem [{\citenamefont {Harland-Lang}\ \emph {et~al.}(2016)\citenamefont
  {Harland-Lang}, \citenamefont {Khoze},\ and\ \citenamefont
  {Ryskin}}]{Harland-Lang:2016qjy}%
  \BibitemOpen
  \bibfield  {author} {\bibinfo {author} {\bibfnamefont {L.~A.}\ \bibnamefont
  {Harland-Lang}}, \bibinfo {author} {\bibfnamefont {V.~A.}\ \bibnamefont
  {Khoze}}, \ and\ \bibinfo {author} {\bibfnamefont {M.~G.}\ \bibnamefont
  {Ryskin}},\ }\href@noop {} {\  (\bibinfo {year} {2016})},\ \Eprint
  {http://arxiv.org/abs/1601.07187} {arXiv:1601.07187 [hep-ph]} \BibitemShut
  {NoStop}%
\bibitem [{\citenamefont {Fichet}\ \emph {et~al.}(2014)\citenamefont {Fichet},
  \citenamefont {von Gersdorff}, \citenamefont {Kepka}, \citenamefont {Lenzi},
  \citenamefont {Royon},\ and\ \citenamefont {Saimpert}}]{Fichet:2013gsa}%
  \BibitemOpen
  \bibfield  {author} {\bibinfo {author} {\bibfnamefont {S.}~\bibnamefont
  {Fichet}}, \bibinfo {author} {\bibfnamefont {G.}~\bibnamefont {von
  Gersdorff}}, \bibinfo {author} {\bibfnamefont {O.}~\bibnamefont {Kepka}},
  \bibinfo {author} {\bibfnamefont {B.}~\bibnamefont {Lenzi}}, \bibinfo
  {author} {\bibfnamefont {C.}~\bibnamefont {Royon}}, \ and\ \bibinfo {author}
  {\bibfnamefont {M.}~\bibnamefont {Saimpert}},\ }\href {\doibase
  10.1103/PhysRevD.89.114004} {\bibfield  {journal} {\bibinfo  {journal} {Phys.
  Rev.}\ }\textbf {\bibinfo {volume} {D89}},\ \bibinfo {pages} {114004}
  (\bibinfo {year} {2014})},\ \Eprint {http://arxiv.org/abs/1312.5153}
  {arXiv:1312.5153 [hep-ph]} \BibitemShut {NoStop}%
\bibitem [{\citenamefont {Dittmaier}\ \emph {et~al.}(2011)\citenamefont
  {Dittmaier} \emph {et~al.}}]{Dittmaier:2011ti}%
  \BibitemOpen
  \bibfield  {author} {\bibinfo {author} {\bibfnamefont {S.}~\bibnamefont
  {Dittmaier}} \emph {et~al.} (\bibinfo {collaboration} {LHC Higgs Cross
  Section Working Group}),\ }\href {\doibase 10.5170/CERN-2011-002} {\
  (\bibinfo {year} {2011}),\ 10.5170/CERN-2011-002},\ \Eprint
  {http://arxiv.org/abs/1101.0593} {arXiv:1101.0593 [hep-ph]} \BibitemShut
  {NoStop}%
\bibitem [{\citenamefont {Aad}\ \emph {et~al.}(2014)\citenamefont {Aad} \emph
  {et~al.}}]{Aad:2014fha}%
  \BibitemOpen
  \bibfield  {author} {\bibinfo {author} {\bibfnamefont {G.}~\bibnamefont
  {Aad}} \emph {et~al.} (\bibinfo {collaboration} {ATLAS}),\ }\href {\doibase
  10.1016/j.physletb.2014.10.002} {\bibfield  {journal} {\bibinfo  {journal}
  {Phys. Lett.}\ }\textbf {\bibinfo {volume} {B738}},\ \bibinfo {pages} {428}
  (\bibinfo {year} {2014})},\ \Eprint {http://arxiv.org/abs/1407.8150}
  {arXiv:1407.8150 [hep-ex]} \BibitemShut {NoStop}%
\bibitem [{\citenamefont {Aad}\ \emph {et~al.}(2016{\natexlab{a}})\citenamefont
  {Aad} \emph {et~al.}}]{Aad:2015kna}%
  \BibitemOpen
  \bibfield  {author} {\bibinfo {author} {\bibfnamefont {G.}~\bibnamefont
  {Aad}} \emph {et~al.} (\bibinfo {collaboration} {ATLAS}),\ }\href {\doibase
  10.1140/epjc/s10052-015-3820-z} {\bibfield  {journal} {\bibinfo  {journal}
  {Eur. Phys. J.}\ }\textbf {\bibinfo {volume} {C76}},\ \bibinfo {pages} {45}
  (\bibinfo {year} {2016}{\natexlab{a}})},\ \Eprint
  {http://arxiv.org/abs/1507.05930} {arXiv:1507.05930 [hep-ex]} \BibitemShut
  {NoStop}%
\bibitem [{\citenamefont {Khachatryan}\ \emph {et~al.}(2015)\citenamefont
  {Khachatryan} \emph {et~al.}}]{Khachatryan:2015cwa}%
  \BibitemOpen
  \bibfield  {author} {\bibinfo {author} {\bibfnamefont {V.}~\bibnamefont
  {Khachatryan}} \emph {et~al.} (\bibinfo {collaboration} {CMS}),\ }\href
  {\doibase 10.1007/JHEP10(2015)144} {\bibfield  {journal} {\bibinfo  {journal}
  {JHEP}\ }\textbf {\bibinfo {volume} {10}},\ \bibinfo {pages} {144} (\bibinfo
  {year} {2015})},\ \Eprint {http://arxiv.org/abs/1504.00936} {arXiv:1504.00936
  [hep-ex]} \BibitemShut {NoStop}%
\bibitem [{\citenamefont {Aad}\ \emph {et~al.}(2016{\natexlab{b}})\citenamefont
  {Aad} \emph {et~al.}}]{Aad:2015agg}%
  \BibitemOpen
  \bibfield  {author} {\bibinfo {author} {\bibfnamefont {G.}~\bibnamefont
  {Aad}} \emph {et~al.} (\bibinfo {collaboration} {ATLAS}),\ }\href {\doibase
  10.1007/JHEP01(2016)032} {\bibfield  {journal} {\bibinfo  {journal} {JHEP}\
  }\textbf {\bibinfo {volume} {01}},\ \bibinfo {pages} {032} (\bibinfo {year}
  {2016}{\natexlab{b}})},\ \Eprint {http://arxiv.org/abs/1509.00389}
  {arXiv:1509.00389 [hep-ex]} \BibitemShut {NoStop}%
\bibitem [{Note3()}]{Note3}%
  \BibitemOpen
  \bibinfo {note} {This can be also true for inelastic electroweak boson
  fusion, to the extent that the $WW$, $ZZ$, $Z\gamma $ fusion diagrams are
  dominated by inelastic photon fusion, which typically occurs, and provided
  that the contribution of the $\phi |D^\mu H|$ operator is
  negligible.}\BibitemShut {Stop}%
\bibitem [{\citenamefont {Chapon}\ \emph {et~al.}(2010)\citenamefont {Chapon},
  \citenamefont {Royon},\ and\ \citenamefont {Kepka}}]{Chapon:2009hh}%
  \BibitemOpen
  \bibfield  {author} {\bibinfo {author} {\bibfnamefont {E.}~\bibnamefont
  {Chapon}}, \bibinfo {author} {\bibfnamefont {C.}~\bibnamefont {Royon}}, \
  and\ \bibinfo {author} {\bibfnamefont {O.}~\bibnamefont {Kepka}},\ }\href
  {\doibase 10.1103/PhysRevD.81.074003} {\bibfield  {journal} {\bibinfo
  {journal} {Phys. Rev.}\ }\textbf {\bibinfo {volume} {D81}},\ \bibinfo {pages}
  {074003} (\bibinfo {year} {2010})},\ \Eprint {http://arxiv.org/abs/0912.5161}
  {arXiv:0912.5161 [hep-ph]} \BibitemShut {NoStop}%
\bibitem [{\citenamefont {Kepka}\ and\ \citenamefont
  {Royon}(2008)}]{Kepka:2008yx}%
  \BibitemOpen
  \bibfield  {author} {\bibinfo {author} {\bibfnamefont {O.}~\bibnamefont
  {Kepka}}\ and\ \bibinfo {author} {\bibfnamefont {C.}~\bibnamefont {Royon}},\
  }\href {\doibase 10.1103/PhysRevD.78.073005} {\bibfield  {journal} {\bibinfo
  {journal} {Phys. Rev.}\ }\textbf {\bibinfo {volume} {D78}},\ \bibinfo {pages}
  {073005} (\bibinfo {year} {2008})},\ \Eprint {http://arxiv.org/abs/0808.0322}
  {arXiv:0808.0322 [hep-ph]} \BibitemShut {NoStop}%
\end{thebibliography}%

\end{document}